\documentclass[sigconf]{acmart}
\usepackage{multirow}
\usepackage{adjustbox}
\usepackage{enumitem}
\usepackage{ragged2e}
\usepackage{array}
\usepackage{xspace}
\usepackage{placeins}
\usepackage{float}
\usepackage{dblfloatfix}
\newcommand{\method}{\textsc{TWISTER}\xspace}
\newcommand{\qw}[1]{#1}

\AtBeginDocument{%
  }

\setcopyright{acmlicensed}
\copyrightyear{2018}
\acmYear{2018}
\acmDOI{XXXXXXX.XXXXXXX}
\acmConference[Conference]{}{June 03--05,
  2018}{Woodstock, NY}
\acmISBN{978-1-4503-XXXX-X/2018/06}

\begin{document}

\title{When Reviews Are Scarce: Structure-Aware Review Imputation for Recommendation in Realistic Sparse Review Settings}


\author{Leyao Wang}
\authornote{These authors contributed equally to this work.}
\affiliation{\institution{Yale University}
\country{}}
\email{leyao.wang.lw855@yale.edu}

\author{Xutao Mao}
\authornotemark[1]
\affiliation{\institution{Vanderbilt University}
\country{}}
\email{xutao.mao@vanderbilt.edu}

\author{Qinwen Ge}
\authornotemark[1]
\affiliation{\institution{Vanderbilt University}
\country{}}
\email{qinwen.ge@vanderbilt.edu}

\author{Xuhui Zhan}
\affiliation{\institution{Vanderbilt University}
\country{}}
\email{xuhui.zhan@vanderbilt.edu}

\author{Yuying Zhao}
\affiliation{\institution{Vanderbilt University}
\country{}}
\email{yuying.zhao@vanderbilt.edu}

\author{Bo Ni}
\affiliation{\institution{Vanderbilt University}
\country{}}
\email{bo.ni@vanderbilt.edu}

\author{Ryan A. Rossi}
\affiliation{\institution{Adobe Research}
\country{}}
\email{ryrossi@adobe.com}

\author{Nesreen K. Ahmed}
\affiliation{\institution{Cisco AI Research}
\country{}}
\email{nesahmed@cisco.com}

\author{Tyler Derr}
\affiliation{\institution{Vanderbilt University}
\country{}}
\email{tyler.derr@vanderbilt.edu}

\begin{abstract}
Textual reviews enrich recommender systems with fine-grained preference signals and enhanced explainability. However, in real-world scenarios, users rarely leave reviews, 
resulting in severe sparsity that undermines the effectiveness of review-aware recommenders under partial observability.
A natural remedy is to synthesize the missing reviews to restore the textual signal these recommenders rely on.
Yet, conventional imputation techniques---such as matrix completion and LLM-based augmentation---either lose contextualized semantics by embedding texts into vectors, or overlook structural dependencies on the user-item graph. 
To address these shortcomings, we propose \textbf{\method}~(\textbf{\underline{T}}o\textbf{\underline{W}}ards \textbf{\underline{I}}mputation on \textbf{\underline{S}}parsity with \textbf{\underline{T}}extual \textbf{\underline{E}}dge Graph \textbf{\underline{R}}epresentation), 
a unified framework for sparse-text recommendation that synthesizes missing reviews by jointly modeling semantic and structural signals.
Specifically, we represent user-item interactions as a Textual-Edge Graph (TEG), treating reviews as edge attributes. To capture relational context, we construct line-graph views and employ an LLM as a graph-aware aggregator. For each interaction lacking a textual review, our model aggregates the neighborhood’s natural-language representations to generate a coherent and personalized review.
Experiments on multiple datasets show that \method{} consistently outperforms traditional numeric, graph-based, and LLM baselines, delivering higher-quality imputed reviews and, more importantly, improved downstream recommendation. In summary, \method{}\footnote{Our code is available at: \url{https://anonymous.4open.science/r/TWISTER-1706}} is a drop-in, structure-aware imputer that recovers the textual signal that review-aware recommenders need when reviews are scarce.
\vspace{-2ex}
\end{abstract}




\maketitle

\begin{figure}[t!]
    \centering
    \includegraphics[width=\linewidth]{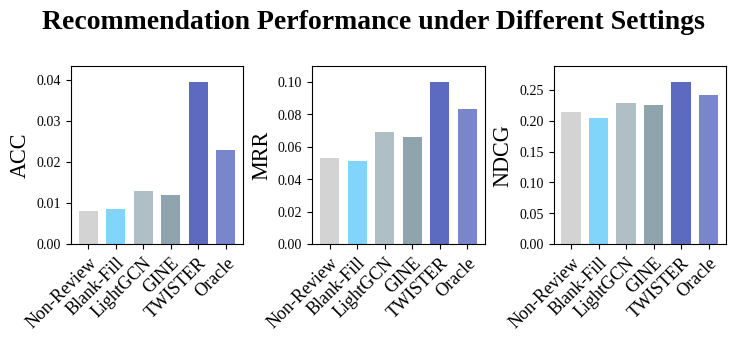}
    \vskip -1em
   \caption{
Recommendation performance averaged over \texttt{Amazon\_Video\_Games} and
\texttt{Goodreads\_Children}. We compare \texttt{Non-Review}, \texttt{Blank-Fill},
interaction-only \texttt{LightGCN}, rating-only \texttt{GINE} (GINE-Rating),
\texttt{TWISTER} (TWISTER-Q), and an \texttt{Oracle} with ground-truth reviews
for all ratings. TWISTER achieves the strongest performance on all three
reported metrics, motivating structure-aware review imputation.
}
     \vskip -1em
    \label{fig:motivation}
\end{figure}

\section{Introduction}

Review-aware recommendation systems \cite{10.1145/3744661,hasan2025reviewbasedrecommendersystemssurvey,jin2023edgeformers,10.1145/3308560.3316601,10.1145/3580305.3599502,10.1145/2806416.2806504} have gained increasing attention due to their ability to leverage textual reviews in addition to user-item interactions or ratings. These reviews not only provide richer semantic signals but also improve model explainability and enable more personalized recommendations. However, in real-world settings, review sparsity is a major obstacle: many users do not leave reviews, especially in cold-start scenarios involving new users or items \cite{lei2025feature,vartak2017meta}. Most existing models assume reviews are available for every interaction, leading to degraded performance when applied under partial observability. As shown in Figure~\ref{fig:motivation}, when only 50\% of ratings
are accompanied by reviews, simply discarding or blank-filling the
missing text degrades recommendation performance. Moreover, interaction-only \texttt{LightGCN} and
\texttt{GINE-Rating}, which respectively rely on interactions alone
and graph structure with ratings but no review text, still leave
substantial room for improvement. These observations motivate
recovering useful textual signals for recommendation under sparse
review text.

A natural approach is therefore to recover the missing textual signal
through review imputation.
Traditional statistical techniques such as Matrix Completion \cite{kuleshov2015tensorfactorizationmatrixfactorization} and deep learning methods like Autoencoders \cite{bank2021autoencoders} have been extensively explored. However, these approaches are primarily designed for numeric or categorical data and do not directly support natural language imputation; a straightforward solution is to embed textual data and perform imputation in the feature space, but this often results in semantic loss \cite{wang2025savetagsemanticawarevicinalrisk,huang-etal-2021-disentangling,sun2025textembeddingscaptureimplicit}. Recent work on large language model (LLM)-based data augmentation \cite{zhou2024surveydataaugmentationlarge, ding2024dataaugmentationusinglarge,wang2025savetagsemanticawarevicinalrisk} attempts to mitigate this by generating novel textual data informed by external knowledge. While promising, these methods still suffer from a key limitation shared with prior imputation techniques: they typically ignore the underlying relational structure of the data, treating features independently or relying solely on generic similarity. In response, graph-based imputation methods such as GRAPE \cite{you2020handlingmissingdatagraph} and Variational Graph Auto-Encoders (VGAE) \cite{kipf2016variationalgraphautoencoders} model the structure among samples and enable information propagation across the graph. These approaches capture local dependencies and structural contexts, but are limited to numerical or categorical attributes. For instance, GRAPE \cite{you2020handlingmissingdatagraph} frames imputation as edge label prediction using GNNs, where edge attributes must be numeric—leading again to potential semantic loss when applied to textual data \cite{wang2025savetagsemanticawarevicinalrisk,huang-etal-2021-disentangling,sun2025textembeddingscaptureimplicit}. 

As a result, existing methods either (i) discard the richness of natural language by embedding texts into dense vectors, or (ii) fail to model relational and structural dependencies among interactions. To overcome these challenges, our work aim to \emph{jointly} model graph topology and textual content to improve recommendation under sparse review text; we do this via structure-aware review synthesis.
Accordingly, we study recommendation under sparse review text and approach it via structure-aware review synthesis on Textual-Edge Graphs (TEGs) \cite{jin2023edgeformers,li2024teg,ling2024linkpredictiontextualedge}, where review texts are naturally represented as edge attributes. 
However, standard graph neural networks (GNNs) are designed for structured, numeric data and cannot directly process raw natural language.
To address this, we leverage large language models (LLMs) as graph-aware imputers. Drawing inspiration from Graph-aware Convolutional LLMs (GaCLLM) \cite{du2024largelanguagemodelgraph}, which demonstrate that LLMs can emulate the message-passing mechanisms of GNNs, we introduce \textbf{\method}~(\textbf{\underline{T}}o\textbf{\underline{W}}ards  \textbf{\underline{I}}mputation on \textbf{\underline{S}}parsity with \textbf{\underline{T}}extual \textbf{\underline{E}}dge Graph \textbf{\underline{R}}epresentation). Our approach first transforms the TEG into its line-graph representation \cite{harary1960some,Zhao_2021}, treating each edge as a node with associated textual information. We then use an LLM to aggregate contextual information from neighboring edges in the line graph and generate the missing review texts. The synthesized reviews are then fed back to review-aware recommenders to improve recommendation under sparsity. This unified framework preserves both the semantic richness of natural language and the structural relationships captured by the graph.

We validate our framework through extensive experiments across three dimensions: recommendation utility, generated text quality, and structural smoothness. Our model consistently generates coherent, personalized reviews aligned with user-item context, and improves downstream recommendation performance.

To summarize, we make the following key contributions toward 
better recommendations under a real-world sparse review setting:
\begin{itemize}[leftmargin=2em,topsep=1.5pt,itemsep=1pt]
    \item We propose \method{}, a structure-aware review imputer using an LLM as a graph aggregator on user/item-side line-graph views, preserving semantic fidelity and structural coherence.
    \item We analyze why this helps: a Dirichlet energy on the interaction graph controls recommender generalization, and \method{} lands the completed reviews in an intermediate-smoothness regime.
    \item We evaluate \method{} along three axes (recommendation utility, review quality, structural smoothness) across five Amazon and Goodreads benchmarks, with consistent gains.
\end{itemize}

The remainder of this paper is organized as follows. Section~\ref{sec:problem} formalizes the problem of recommendation under sparse review text. Section~\ref{sec:method} introduces \method{}, our LLM-based structure-aware review imputer. Section~\ref{sec:exp} reports empirical results on recommendation utility and review quality, and Section~\ref{sec:theory} develops a Dirichlet-energy analysis that explains these gains. Section~\ref{sec:related} discusses related work, and Section~\ref{sec:conclusion} concludes.

\vspace{-1em}

\section{Problem Statement}\label{sec:problem}
\begingroup
\setlength\abovedisplayskip{3pt}
\setlength\belowdisplayskip{3pt}
\setlength\abovedisplayshortskip{2pt}
\setlength\belowdisplayshortskip{2pt}

Following standard notation~\cite{hasan2025reviewbasedrecommendersystemssurvey}, let the user and item sets be $\mathcal{U}=\{u_1,\dots,u_{n}\}$ and $\mathcal{I}=\{i_1,\dots,i_{m}\}$ with $n=|\mathcal{U}|$ and $m=|\mathcal{I}|$. Each item $i \in \mathcal{I}$ is associated with structured metadata $s_{i}$. Scalar ratings are stored in a matrix $\mathbf{Y}=[y_{ui}]\in\mathbb{R}^{n\times m}$, where the interaction set $\Gamma=\{(u,i)\mid y_{ui}\ \text{observed}\}\subseteq\mathcal{U}\times\mathcal{I}$ indexes available interactions.

Each observed user--item pair is optionally accompanied by a free-form textual review; we collect all reviews in $\mathbf{R}=[r_{ui}]\in\mathcal{T}^{n\times m}$, where $r_{ui}=\varnothing$ if no review is written. We define the mask $\mathbf{M}\in\{0,1\}^{n\times m}$ with $\mathbf{M}_{ui}=1 \iff r_{ui}=\varnothing$, and write $\Omega = \{(u,i) \in \Gamma \mid \mathbf{M}_{ui}=1\}$ for the index set of text-missing pairs.

Finally, let $\mathsf{T}$ denote a downstream recommendation task with output space $\mathcal{O}_{\mathsf{T}}$ and evaluation metric $\mathcal{S}_{\mathsf{T}}$. Given $(\mathbf{Y},\mathbf{R},\mathbf{M},\Gamma)$, the goal is to learn a review-aware recommender
\[ 
f_{\mathsf{T}}:\;(\mathcal{U},\mathcal{I},\mathbf{Y},\mathbf{R})\;\longrightarrow\;\mathcal{O}_{\mathsf{T}}.
\]

Real-world recommender systems collect abundant implicit feedback in the form of user--item interactions, but the explicit feedback most useful for review-aware recommendation, written reviews, is sparse: most users rate without writing, and even when they do, their reviews cover only a small fraction of their interactions. In practice this means $\mathbf{Y}$ is densely observed while $\mathbf{R}$ is largely missing, with $|\Omega|$ typically covering the majority of $\Gamma$.

\medskip\noindent\textbf{Objective.}\;
Given $(\mathbf{Y}, \mathbf{R}, \mathbf{M}, \Gamma)$, learn a review-aware recommender $f_{\mathsf{T}}$ that maximizes the downstream metric $\mathcal{S}_{\mathsf{T}}$ despite the sparsity of $\mathbf{R}$. As Figure~\ref{fig:motivation} shows, naive responses (discarding $\mathbf{R}$ or blank-filling the missing entries) underperform even the text-free baseline: solving this problem requires recovering the missing review text before training $f_{\mathsf{T}}$. In §\ref{sec:method}, we propose \method{}, an LLM-based review imputer that addresses this problem.

\endgroup

\section{TWISTER: Methodology}
\label{sec:method}

Review-aware recommenders require textual signal on every interaction, yet in practice most $(u,i) \in \Gamma$ carry no review ($\mathbf{M}_{ui}=1$). Our goal is to learn an imputer $\Psi_\theta$ that supplies, for every such pair, a review $\hat{r}_{ui}$ plausible enough that a downstream recommender trained on the completed matrix $\widehat{\mathbf{R}}$ recovers the utility it would have under full observation. Two design requirements follow directly from this goal: each imputed review must be (i) \emph{grounded} in the specific $(u,i)$ pair it describes, matching the user's writing style and the item's attributes, and (ii) \emph{informed by relational context}, drawing on evidence from other interactions that share an endpoint with $(u,i)$. \method{} meets both requirements through the pipeline illustrated in Figure 2: we represent user--item interactions as a Textual-Edge Graph (TEG), carving user/item-side line-graph views over this TEG, and using an LLM as a graph-aware aggregator that condenses each line-node's neighborhood into textual context for generation. This construction also admits a Dirichlet-energy analysis of why the resulting completions help downstream recommenders, which we develop in \S\ref{sec:theory}.

\subsection{Sparse Review Imputation: Formalization}
\label{sec:formalization}

Building on the notation of \S\ref{sec:problem}, where $\Omega = \{(u,i) \in \Gamma \mid \mathbf{M}_{ui}=1\}$ indexes the user--item pairs whose ratings are observed but whose review texts are absent, our goal is to endow every $(u,i) \in \Omega$ with a plausible, context-aware review. We frame this as learning an imputer (text generator)
\(
  \Psi_{\boldsymbol{\theta}} : \mathcal{P}\rightarrow\mathcal{T},
\)
where each prompt \(p_{ui}\in\mathcal{P}\) may gather the following: \textbf{(i)} numeric rating $y_{ui}$; ~ \textbf{(ii)} item metadata $s_i$; ~ \textbf{(iii)} all observed reviews $\mathcal{R}_{\mathrm{obs}} := \{\, r_{u',i'} \mid (u',i') \in \Gamma,\; \mathbf{M}_{u'i'} = 0 \}$.

\vspace{-1.5ex}
\medskip\noindent\textbf{Imputed review matrix.}\;
The completed review matrix is 
\[
\widehat r_{ui}=
  \begin{cases}
    r_{ui}, &\mathbf  M_{ui}=0,\\
    \Psi_{\boldsymbol{\theta}}(p_{ui}), &\mathbf   M_{ui}=1,
  \end{cases}
\qquad
\widehat{\mathbf R}=[\widehat r_{ui}].
\]

\vspace{-1.5ex}
\medskip\noindent\textbf{Learning objective.}\; For downstream task $\mathsf T$ with metric $\mathcal S_{\mathsf T}(\cdot)$, 
we choose parameters $\boldsymbol{\theta}$ for imputer $\Psi_{\boldsymbol{\theta}}$ as $\boldsymbol{\theta}^\star =\arg\max_{\boldsymbol{\theta}} \mathcal S_{\mathsf T}\hspace{0.25ex}\!\bigl(f_{\mathsf T}(\mathbf Y, \widehat{\mathbf R})\bigr),$

i.e., we seek the imputer that maximizes $\mathsf T$'s performance. Unless noted otherwise, $f_{\mathsf T}$ is trained after $\widehat{\mathbf R}$ is fixed, so improvements in $\mathcal S_{\mathsf T}$ are measured relative to training on the incomplete review matrix without imputation.

\begin{figure}[t!]
    \hspace*{0.75ex}\includegraphics[width=1.05\linewidth]{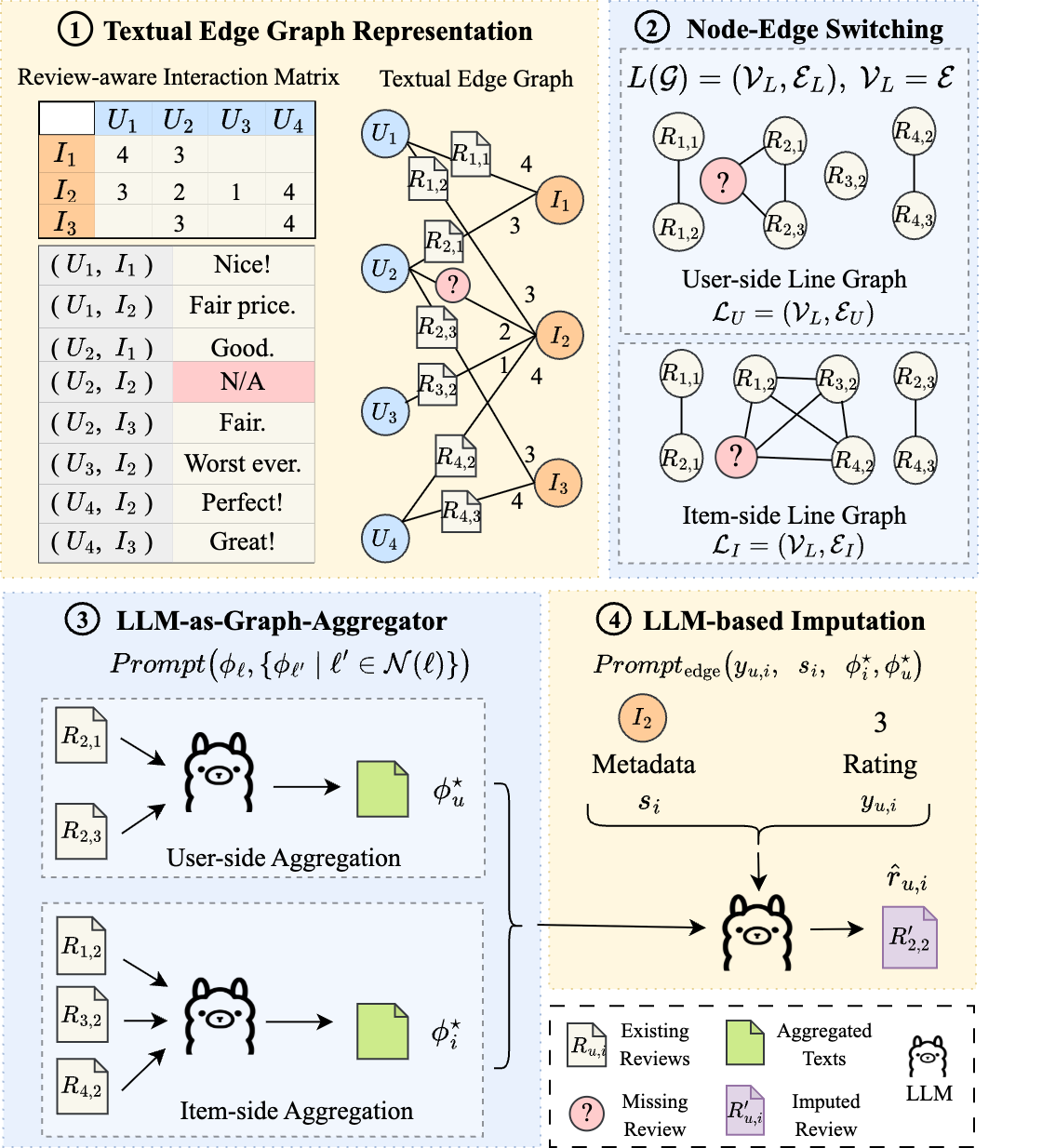}
    \caption{Pipeline for review imputation. (1) Construct a bipartite textual-edge graph from user–item interactions and reviews. (2) Build user-side and item-side line graphs to capture relational context. (3) Use an LLM to aggregate neighborhood information into textual representations. (4) Prompt the LLM with these textual representations, ratings, and metadata to generate missing reviews.}
     \vskip -1ex
    \label{fig:twister}
\end{figure}

\subsection{LLM-based Missing Review Imputation}
\label{sec:edge-impute}
We now fill every user--item pair with a missing review ($\mathbf{M}_{u,i}=1$) by defining an imputer
$\Psi_{\boldsymbol{\theta}} : \mathcal{P} \to \mathcal{T}$,
where $\mathcal{P}$ is the space of natural-language prompts and $\mathcal{T}$ is the space of textual reviews. The prompt draws on two contexts---a user context $\phi^\star_u$ and an item context $\phi^\star_i$---whose construction from the neighborhood of $(u,i)$ we defer to \S\ref{sec:context-extraction}. We instantiate $\Psi_{\boldsymbol{\theta}}$ via prompt-based generation:

\paragraph{Prompt construction.}
For each such edge we assemble a natural-language prompt
\(p_{u,i}\in\mathcal{P}\) that contains four ingredients:

\begin{enumerate}[leftmargin=2em,itemsep=1pt]
\item \textbf{Rating cue.}  
      The (rescaled) numeric rating
      \(y_{u,i}\), 
\item \textbf{Item metadata.} The product description $s_i$,

\item \textbf{Item context.}  
      The item representations \(\phi_{i}^{\star}\)
     from aggregation,
      optionally followed by structured metadata $s_i$.
\item \textbf{User context.}  
      The user representations \(\phi_{u}^{\star}\),
      conveying writing style and preference profile.
\end{enumerate}
\begin{equation}
p_{u,i} \;=\; \textsc{Prompt}_{\text{edge}}\bigl(y_{u,i},\; s_i,\; \phi_{i}^{\star},\; \phi_{u}^{\star}\bigr).
\label{eq:edge-prompt}
\end{equation}

\paragraph{Generation.} A pretrained backbone LLM is queried with Eq.~\eqref{eq:edge-prompt} to produce the imputed review:
\begin{equation}
\widehat{r}_{u,i} = \Psi_{\boldsymbol{\theta}}(p_{u,i}) = \mathrm{LLM}(p_{u,i}), \quad (u,i) \in \Omega.
\label{eq:edge-gen}
\end{equation}
Plugging $\widehat{r}_{u,i}$ back into the completed review matrix $\widehat{\mathbf{R}}$ defined in \S\ref{sec:formalization} yields a fully review-annotated TEG on which downstream recommenders can be trained (\S\ref{sec:downstream}).

\subsection{User and Item Context Extraction}
\label{sec:context-extraction}

To obtain the contextual representations $\phi^\star_u$ and $\phi^\star_i$ left open in \S\ref{sec:edge-impute}, we extract them from the local neighborhood of each text-missing edge $(u,i) \in \Omega$ in three steps. First, we represent the review-aware interaction matrix as a bipartite Textual-Edge Graph (TEG) that attaches rating--review payloads directly to edges (\S\ref{sec:bipartite-teg}). Second, to aggregate local information around each missing edge, we convert the TEG into line-graph views---each interaction becomes a node, structurally related interactions become neighbors---and propose three variants that focus attention on the user side, the item side, and a weighted user-side accounting for item--item similarity (\S\ref{sec:line-graphs}). Third, we deploy an LLM as a graph-aware aggregator that condenses each line-node's textual neighborhood into $\phi^\star_u$ and $\phi^\star_i$ (\S\ref{sec:llm-conv-full}).

\subsubsection{Review-Aware Interaction Matrix as a Bipartite TEG}
\label{sec:bipartite-teg}

A textual-edge graph (TEG) \cite{li2024teg,jin2023edgeformers,ling2024linkpredictiontextualedge} is an undirected graph whose edges carry textual payloads. We instantiate a bipartite TEG $\mathcal{G} = (\mathcal{V}, \mathcal{E}_\phi)$ with $\mathcal{V} = \mathcal{U} \cup \mathcal{I}$ and edge set $\mathcal{E}_\phi = \{(u, i, \phi_{u,i}) \mid (u,i) \in \Gamma\}$, where each payload pairs rating with review: $\phi_{u,i} = [y_{u,i};\, r_{u,i}]$ when $\mathbf{M}_{u,i}=0$ and $\phi_{u,i} = [y_{u,i};\, \varnothing]$ when $\mathbf{M}_{u,i}=1$. This integrates structural and textual information directly into the edge set, eliminating the need for a separate review collection.

\subsubsection{Conversion to Line Graphs}
\label{sec:line-graphs}
Starting from the bipartite TEG
\(
  \mathcal{G}= (\mathcal{V},\mathcal{E}_\phi)
\)
defined in §\ref{sec:bipartite-teg},
each interaction edge
\(e=(u,i,\boldsymbol{\phi}_{u,i})\in\mathcal{E}_\phi\)
is “reified’’ into a \emph{line-node}.
Formally, the line graph of $\mathcal{G}$ is 
\vspace{-2pt}
\[
  L(\mathcal{G}) = (\mathcal{V}_L,\mathcal{E}_L),
  \qquad
  \mathcal{V}_L = \mathcal{E}_\phi,
\]

\vspace{-2pt}\noindent
so that every $\ell_e\in\mathcal{V}_L$ corresponds to one
user–item interaction.
Two line-nodes $\ell_{e_1},\ell_{e_2}$ are adjacent in
$\mathcal{E}_L$ precisely when the underlying edges share an endpoint
in~$\mathcal{G}$.

\vspace{2pt}\noindent
\textbf{Context-specific line-graph views.}
We carve three task-oriented sub-graphs from $L(\mathcal{G})$:

\begin{enumerate}[leftmargin=*,itemsep=4pt]
  \item \textbf{User-side line graph}
        \(\mathcal{L}_{ U}=(\mathcal{V}_L,\mathcal{E}_{ U})\).  
        An edge \(\{\ell_{e_1},\ell_{e_2}\}\in\mathcal{E}_{ U}\)
        exists if and only if both interactions involve the \emph{same user}
        \(u\), i.e.\ \(e_1=(u,i_1,\cdot)\) and \(e_2=(u,i_2,\cdot)\).
        All edges are unweighted.  
        
        \emph{Captures:} a single user’s multi-item context.

  \item \textbf{Item-side line graph}
        \(\mathcal{L}_{I}=(\mathcal{V}_L,\mathcal{E}_{I})\).  
        \(\{\ell_{e_1},\ell_{e_2}\}\in\mathcal{E}_{I}\)
        if and only if both interactions share the \emph{same item}
        \(i\).  
        
        \emph{Captures:} crowd opinions focused on one item.

  \item \textbf{Weighted user-side line graph}
        \(\mathcal{L}_{U,w}=(\mathcal{V}_L,\mathcal{E}_{U},W)\).  
        The topology equals $\mathcal{L}_{U}$, but each edge carries
        \(
          W_{\ell_{e_1}\ell_{e_2}}
          = \operatorname{sim}\bigl(
              i_{e_1},\,i_{e_2}
            \bigr),
        \)
        where \(\operatorname{sim}\) is cosine similarity between
        item-text encodings. 
        
        \emph{Captures:} fine-grained preference clusters—two
        interactions by the same user are “closer’’ when 
        items
        are semantically similar.
\end{enumerate}
Each view yields a Laplacian matrix
\(
  \mathbf{L}_{U},\;
  \mathbf{L}_{I},\;
  \mathbf{L}_{U,w}
\),
which can be used when computing our energy objective in $\S$ \ref{sec:dirichlet}.

\subsubsection{LLM as Graph Aggregator}
\label{sec:llm-conv-full}
Having represented each user–item interaction as a line-node
$\ell_{(u,i)} \in \mathcal{V}_{L}$ (§\ref{sec:line-graphs}), we now require a mechanism that can \emph{jointly reason over edge text and relational structure}. Rather than relying on hand-crafted GNN layers, we follow \citet{du2024largelanguagemodelgraph} and employ
language-model aggregation directly on the \textit{line graph}.

We continue to utilize the context-specific line-graph views and associated notations as defined in $\S$\ref{sec:line-graphs}. For any line-node $\ell$, we denote its textual payload (review $+$ rating) by $\phi_{\ell}$. We then formalize both \textit{user-side aggregation} and \textit{item-side aggregation}, corresponding to the user-side and item-side line graphs, respectively.

\paragraph{User-side Aggregation}

The neighbors of a line-node $\ell$ in the user-side line graph $\mathcal{L}_{U}=(\mathcal{V}_L,\mathcal{E}_{U})$ are given by
\(
  \mathcal{N}_U(\ell) =
  \{\,\ell' \mid \{\ell,\ell'\}\in\mathcal{E}_{U}\,\}.
\)
We define a single round of user-side aggregation:

\vspace{-1pt}
\begin{equation}
\phi_{U}^{\star} \;=\; \operatorname{LLM}\Bigl(\textsc{Prompt}\bigl(\phi_{\ell},\, \{\phi_{\ell'} \mid \ell' \in \mathcal{N}_{U}(\ell)\}\bigr)\Bigr).
\label{eq:llm-user-agg}
\end{equation}

\vspace{-12pt}
\paragraph{Item-side Aggregation}
Similarly, the neighbors of a line-node $\ell$ in the item-side line graph $\mathcal{L}_{I}=(\mathcal{V}_L,\mathcal{E}_{I})$ are defined by
\(
  \mathcal{N}_I(\ell) =
  \{\,\ell' \mid \{\ell,\ell'\}\in\mathcal{E}_{I}\,\}.
\)
Denote a single round of item-side aggregation:
\vspace{-3pt}
\begin{equation}
\phi_{I}^{\star} \;=\; \operatorname{LLM}\Bigl(\textsc{Prompt}\bigl(\phi_{\ell},\, \{\phi_{\ell'} \mid \ell' \in \mathcal{N}_{I}(\ell)\}\bigr)\Bigr).
\label{eq:llm-item-agg}
\end{equation}

\vspace{-11pt}

\paragraph{Outcome.}  After aggregation, each interaction node $\ell$ retains an LLM-condensed summary of the textual context from structurally related interactions, ready
to further processed for
imputation tasks.

\vspace{9pt}
\noindent

\textbf{Why One-Hop Suffices.}

Our restriction to one-hop neighbors reflects the nature of review generation, which differs from how shallow aggregation is usually justified in recommendation. Prior work (item--item CF~\cite{sarwar2001item}, LightGCN~\cite{he2020lightgcn}, UltraGCN~\cite{mao2021ultragcn}) limits propagation depth to prevent over-smoothing of preference signals that are themselves the recommendation target. Here the aggregator does not produce a ranking score; it generates a textual review that must be grounded in one specific $(u, i)$ pair. Pulling in multi-hop reviews (e.g., reviews by users who share items with users who share items with $u$) dilutes $u$'s first-person writing style and the attribute-grounded language appropriate to $i$ with content from weakly related interactions, hurting fidelity and specificity. Restricting aggregation to one hop in the line graph --- $u$'s other reviews (via $\mathcal{L}_U$) and other users' reviews of $i$ (via $\mathcal{L}_I$) --- keeps generation localized on exactly the two endpoints whose interaction we are reconstructing. This one-hop restriction applies only to the imputer $\Psi_{\boldsymbol{\theta}}$; once $\widehat{\mathbf{R}}$ is fixed, the 
recommender $f_{\mathsf{T}}$ operates on the full TEG with $\widehat{\mathbf{R}}$ and performs multi-hop message passing as usual.

\subsection{Downstream Recommendation with $\widehat{\mathbf{R}}$.}
\label{sec:downstream}
Once every text-missing edge carries an imputed review, the TEG becomes fully review-annotated, with payload $\widehat{\phi}_{u,i} = [y_{u,i}; \widehat{r}_{u,i}]$ for every $(u,i) \in \Gamma$. We then train a standard review-aware recommender $f_{\mathsf{T}}$ on this completed graph, which encodes each edge's review, propagates over multi-hop neighborhoods on $\mathcal{G}$ (not on the line graph, which was used only by the imputer), and optimizes its usual ranking objective. We use Edgeformers~\cite{jin2023edgeformers} as the primary
downstream backbone (Supplementary Table~\ref{tab:gine_results}). Two properties of this design matter: (i) $\Psi_{\boldsymbol{\theta}}$ and $f_{\mathsf{T}}$ are trained in sequence (not jointly), so any gain in $\mathcal{S}_{\mathsf{T}}$ reflects the quality of $\widehat{\mathbf{R}}$ rather than co-adaptation; (ii) from the recommender's perspective, partial observability is resolved, letting $f_{\mathsf{T}}$ operate in the regime it was designed for.

\section{Experiments}\label{sec:exp}

\begin{table*}[t!]
\centering
\setlength{\tabcolsep}{1.2pt}
\caption{Recommendation performance under the \textit{Uniform Masking} setting ($50\%$ of user–item edges retain their reviews). 
}
\vskip -1em
\begin{adjustbox}{width=1\textwidth}
\begin{tabular}{@{}lccccccccccccccccccccc@{}}
\toprule
        & \multicolumn{4}{c}{\textbf{Amazon\_Video}}
        & \multicolumn{4}{c}{\textbf{Amazon\_Music}}
        & \multicolumn{4}{c}{\textbf{Amazon\_Toys}}
        & \multicolumn{4}{c}{\textbf{Goodreads\_Comics}}
        & \multicolumn{4}{c}{\textbf{Goodreads\_Children}}
        & \textbf{Avg.}\\
\cmidrule(l){2-5} \cmidrule(l){6-9} \cmidrule(l){10-13}
\cmidrule(l){14-17} \cmidrule(l){18-21}
\textbf{Method} & ACC & AUC & MRR & NDCG
       & ACC & AUC & MRR & NDCG
       & ACC & AUC & MRR & NDCG
       & ACC & AUC & MRR & NDCG
       & ACC & AUC & MRR & NDCG & 
       \textbf{Rank}\\ \midrule
Blank  & 0.8$\pm$0.3 & 50.1$\pm$1.4 & 4.7$\pm$0.0 & 20.6$\pm$0.1
       & 0.7$\pm$0.2 & 51.4$\pm$0.4 & 6.2$\pm$0.1 & 22.6$\pm$0.2
       & 1.9$\pm$0.5 & 53.1$\pm$0.2 & 9.3$\pm$0.2 & 26.1$\pm$0.2
       & 1.2$\pm$0.1 & 49.9$\pm$0.4 & 5.8$\pm$0.4 & 21.7$\pm$0.5
       & 0.9$\pm$0.3 & 51.0$\pm$0.1 & 5.5$\pm$0.2 & 20.3$\pm$0.4
       & 12.07 \\
Random & 1.0$\pm$0.0 & 50.4$\pm$2.3 & 5.6$\pm$0.4 & 21.5$\pm$0.6
       & 8.9$\pm$1.8 & 58.0$\pm$0.5 & 12.5$\pm$0.7 & 29.5$\pm$0.4
       & 3.6$\pm$0.9 & 52.9$\pm$1.2 & 10.5$\pm$0.4 & 27.8$\pm$0.2
       & 1.3$\pm$0.1 & 50.2$\pm$0.1 & 6.0$\pm$0.2 & 21.9$\pm$0.2
       & 0.9$\pm$0.3 & 52.0$\pm$0.1 & 6.5$\pm$0.2 & 22.5$\pm$0.2
       &  9.05 \\ 
Mean   & 1.4$\pm$0.2 & 50.7$\pm$0.6 & 5.8$\pm$1.2 & 21.6$\pm$1.0
       & 6.9$\pm$0.2 & 61.4$\pm$0.4 & 14.2$\pm$1.7 & 29.4$\pm$0.1
       & 18.4$\pm$1.3 & 69.5$\pm$1.6 & 26.0$\pm$2.1 & 39.1$\pm$0.3
       & 1.1$\pm$0.2 & 49.8$\pm$0.3 & 5.9$\pm$0.3 & 21.5$\pm$0.2
       & 1.3$\pm$0.1 & 50.6$\pm$0.2 & 6.1$\pm$0.4 & 21.8$\pm$0.3
       &  8.07 \\ \midrule
MF     & 1.0$\pm$0.1 & 51.2$\pm$0.9 & 5.5$\pm$0.1 & 21.4$\pm$0.7
       & 5.8$\pm$0.6 & 60.1$\pm$1.2 & 13.9$\pm$0.0 & 28.5$\pm$1.3
       & 15.6$\pm$1.1 & 65.1$\pm$0.1 & 22.7$\pm$0.8 & 37.9$\pm$1.4
       & 1.4$\pm$0.2 & 51.8$\pm$0.5 & 6.3$\pm$0.2 & 22.1$\pm$0.4
       & 1.2$\pm$0.1 & 50.9$\pm$0.3 & 5.7$\pm$0.1 & 21.2$\pm$0.2
       &  8.35 \\
AE     & 1.3$\pm$0.0 & 50.9$\pm$0.1 & 6.0$\pm$1.1 & 21.9$\pm$0.8
       & 5.8$\pm$0.0 & 62.7$\pm$0.1 & 14.7$\pm$0.8 & 30.8$\pm$0.1
       & 19.6$\pm$2.5 & 70.3$\pm$1.2 & 27.4$\pm$1.3 & 40.2$\pm$0.4
       & 1.7$\pm$0.3 & 52.4$\pm$0.2 & 6.8$\pm$0.5 & 22.9$\pm$0.3
       & 1.5$\pm$0.2 & 51.7$\pm$0.4 & 6.2$\pm$0.3 & 22.1$\pm$0.2
       &  5.60 \\
GAIN   & 3.1$\pm$0.6 & 53.5$\pm$0.0 & 7.0$\pm$0.1 & 22.7$\pm$1.7
       & 6.0$\pm$0.0 & 64.2$\pm$2.6 & 14.8$\pm$1.3 & 30.5$\pm$0.8
       & \underline{25.0$\pm$0.4} & \underline{76.6$\pm$0.2} & \underline{34.9$\pm$0.1} & \underline{46.8$\pm$0.2}
       & 2.2$\pm$0.4 & \underline{54.1$\pm$0.7} &\textbf{ 7.6$\pm$0.3} & \textbf{24.2$\pm$0.4}
       & 1.9$\pm$0.7 & 53.0$\pm$0.1 & 6.8$\pm$0.5 & 22.4$\pm$1.5
       &  3.20 \\ \midrule
KNN    & 1.0$\pm$1.0 & 51.6$\pm$0.8 & 5.8$\pm$1.4 & 21.7$\pm$1.3
       & 3.9$\pm$0.1 & 59.7$\pm$0.8 & 10.1$\pm$0.4 & 23.1$\pm$0.0
       & 2.1$\pm$0.3 & 53.9$\pm$1.3 & 9.7$\pm$0.6 & 26.5$\pm$0.3
       & 1.6$\pm$0.4 & 52.2$\pm$0.3 & 6.2$\pm$0.1 & 22.0$\pm$0.0
       & 1.1$\pm$0.2 & 51.4$\pm$1.4 & 5.3$\pm$0.1 & 20.9$\pm$0.4
       &  9.45 \\
GRAPE  & 1.0$\pm$0.0 & 52.0$\pm$0.1 & 4.7$\pm$0.0 & 20.9$\pm$0.0
       & 7.3$\pm$0.1 & 66.4$\pm$2.4 & 17.0$\pm$2.3 & 32.1$\pm$0.3
       & 10.9$\pm$1.0 & 61.3$\pm$0.8 & 19.2$\pm$1.8 & 35.5$\pm$0.0
       & 1.8$\pm$0.2 & 53.6$\pm$0.6 & 7.1$\pm$0.4 & 23.5$\pm$0.1
       & 1.6$\pm$0.3 & 52.1$\pm$0.2 & 6.4$\pm$0.2 & 22.3$\pm$0.2
       &  5.85 \\
VGAE   & 1.1$\pm$0.4 & 48.6$\pm$1.1 & 3.9$\pm$0.1 & 19.8$\pm$0.7
       & 4.4$\pm$0.7 & 58.5$\pm$0.3 & 10.6$\pm$1.4 & 23.0$\pm$0.2
       & 11.7$\pm$1.5 & 63.4$\pm$0.6 & 21.9$\pm$0.8 & 38.3$\pm$1.7
       & 1.3$\pm$0.3 & 49.7$\pm$0.8 & 5.4$\pm$0.6 & 20.8$\pm$0.4
       & 1.0$\pm$0.2 & 49.2$\pm$0.5 & 4.9$\pm$0.3 & 19.9$\pm$0.3
       & 10.88 \\ \midrule
\qw{\texttt{LLM-L}}  & 1.0$\pm$0.3 & 50.8$\pm$0.7 & 5.5$\pm$0.2 & 21.4$\pm$0.3
       & 7.0$\pm$0.2 & 63.0$\pm$0.8 & 15.1$\pm$0.4 & 30.4$\pm$0.5
       & 21.9$\pm$2.6 & 75.7$\pm$0.3 & 31.6$\pm$1.5 & 44.7$\pm$1.1
       & 1.5$\pm$0.2 & 50.3$\pm$0.6 & 6.0$\pm$0.3 & 21.7$\pm$0.2
       & 1.1$\pm$0.6 & 50.9$\pm$0.1 & 5.9$\pm$0.1 & 21.9$\pm$0.0
       &  7.17 \\
\qw{\texttt{LLM-Q}}   & 2.1$\pm$1.0 & 51.9$\pm$0.8 & 5.7$\pm$0.3 & 21.5$\pm$0.2
       & 5.6$\pm$0.3 & 62.7$\pm$0.4 & 14.6$\pm$0.2 & 29.9$\pm$0.3
       & 5.8$\pm$1.9 & 58.4$\pm$1.5 & 18.7$\pm$1.8 & 35.6$\pm$1.5
       & \underline{2.4$\pm$0.1} & 52.9$\pm$0.2 & 6.4$\pm$0.0 & 22.4$\pm$0.0
       & \textbf{4.0$\pm$0.2} & \textbf{54.5$\pm$0.3} & \textbf{7.8$\pm$0.2} & \textbf{25.6$\pm$0.3}
       &  5.70 \\
\qw{\method-L} & \underline{3.7$\pm$1.2} & \underline{53.7$\pm$0.9} & \underline{11.7$\pm$0.5} & \textbf{28.8$\pm$0.5}
       & \underline{7.0$\pm$1.1} & \underline{68.5$\pm$2.4} & \underline{17.1$\pm$1.7} & \underline{32.6$\pm$1.6}
       & \textbf{28.9$\pm$3.5} & \textbf{80.7$\pm$0.8} & \textbf{40.2$\pm$2.8} & \textbf{51.9$\pm$2.2}
       & 1.8$\pm$0.3 & 52.8$\pm$0.4 & \underline{6.9$\pm$0.2} & 23.1$\pm$0.1
       & \underline{2.3$\pm$0.4} & 52.6$\pm$0.2 & 7.1$\pm$0.3 & 23.8$\pm$0.2
       & \textbf{2.55} \\
\qw{\method-Q} & \textbf{5.9$\pm$0.5} & \textbf{59.7$\pm$1.1} & \textbf{12.5$\pm$0.8} & \underline{27.9$\pm$0.7}
       & \textbf{9.6$\pm$0.1} & \textbf{69.6$\pm$0.2} & \textbf{19.0$\pm$0.3} & \textbf{34.1$\pm$0.3}
       & 6.2$\pm$0.5 & 59.1$\pm$0.1 & 19.1$\pm$0.2 & 34.9$\pm$0.8
       & \textbf{2.6$\pm$0.4} & \textbf{54.2$\pm$0.3} & \underline{7.3$\pm$0.2} & \underline{24.1$\pm$0.2}
       & 2.0$\pm$0.2 & \underline{53.8$\pm$0.1} & \underline{7.5$\pm$0.1} & \underline{24.7$\pm$0.1}
       & \underline{3.05} \\
\bottomrule
\end{tabular}
\end{adjustbox}
\vskip -1ex
\label{tab:rec_rr05}

\end{table*}

In this section, we conduct extensive experiments to evaluate the effectiveness of our proposed method. We begin by assessing how well \method~ handles review sparsity in recommendation tasks. We then analyze the source of its performance gains from both \textit{Semantic} and \textit{Structural} perspectives. Specifically, our experiments are designed to address the following research questions (RQs):

\begin{itemize}[leftmargin=1.5em]
\item \textbf{RQ1:} Can \method~ effectively improve recommendation performance under review sparsity?

\vspace{1ex}
\item \textbf{RQ2:} Do imputed texts from \method~ exhibit richer semantics than those from embedding-based methods?

\end{itemize}

We begin by describing our experimental setup, followed by a detailed analysis addressing the proposed research questions.

\subsection{Experimental Setup}\label{ssec:setup}
\subsubsection{Datasets}\label{sec:datasets}

Following prior work on \emph{Textual-Edge Graphs}~(TEGs) \cite{jin2023edgeformers,ling2024linkpredictiontextualedge,li2024teg},
we evaluate our method on two public benchmarks that pair
user–item interactions with free‐text reviews: \textsc{Amazon} Review 2018~\cite{ni-etal-2019-justifying} and \textsc{Goodreads} Book Graph Dataset ~\cite{wan-etal-2019-fine}. We select five datasets for our experiments:  \texttt{Amazon\_Video}, \texttt{Amazon\_Music}, \texttt{Amazon\_Toys}, \texttt{Goodreads\_Comics}, and \texttt{Goodreads\_Children}. Details are reported in 
Table~\ref{tab:dataset}, Appendix \ref{app_subsec:code}.

\vspace{-0.75ex}
\paragraph{Data splits and masking protocols.}
Edges are partitioned into training, validation, and test sets with a
$70{:}10{:}20$ ratio.  To more accurately model real-world review sparsity, we design the following two experimental scenarios: \textbf{(i) Cold Start:} Select $50\%$ of users at random and mask \emph{all} edges incident to them; and \textbf{ (ii) Uniform Masking:} Randomly mask $50\%$ of all edges.

\vspace{-0.75ex}
\subsubsection{Evaluation Metrics}

To assess the utility of imputed reviews, we follow standard protocols in review-aware recommendation~\cite{jin2023edgeformers,ling2024linkpredictiontextualedge,li2024teg}, adopting four widely used metrics—accuracy (ACC), AUC, MRR, and NDCG—to evaluate top-$k$ ($k=10$) recommendation performance. Each experiment is repeated five times with different fixed seeds, and we report the average results.

To evaluate semantic fidelity (similarity with the ground truth reviews), we compare imputed texts against the original masked reviews using ROUGE$_L$ \cite{lin-2004-rouge} and BERT$_{cos}$ \cite{zhang2020bertscoreevaluatingtextgeneration}.
In addition, following prior works, we employ LLM-as-Judge to deliver human-aligned quality assessments across four key dimensions: Authenticity  \cite{meng2025largelanguagemodelshidden}, Helpfulness \cite{Gamzu2021}, Specificity \cite{zhao2025ai}, and Readability \cite{zhao2025ai}. Implementation details are provided in $\S$\ref{sssec:llmjudge}.

\begin{figure*}[t]
    \centering
    \includegraphics[width=\textwidth]{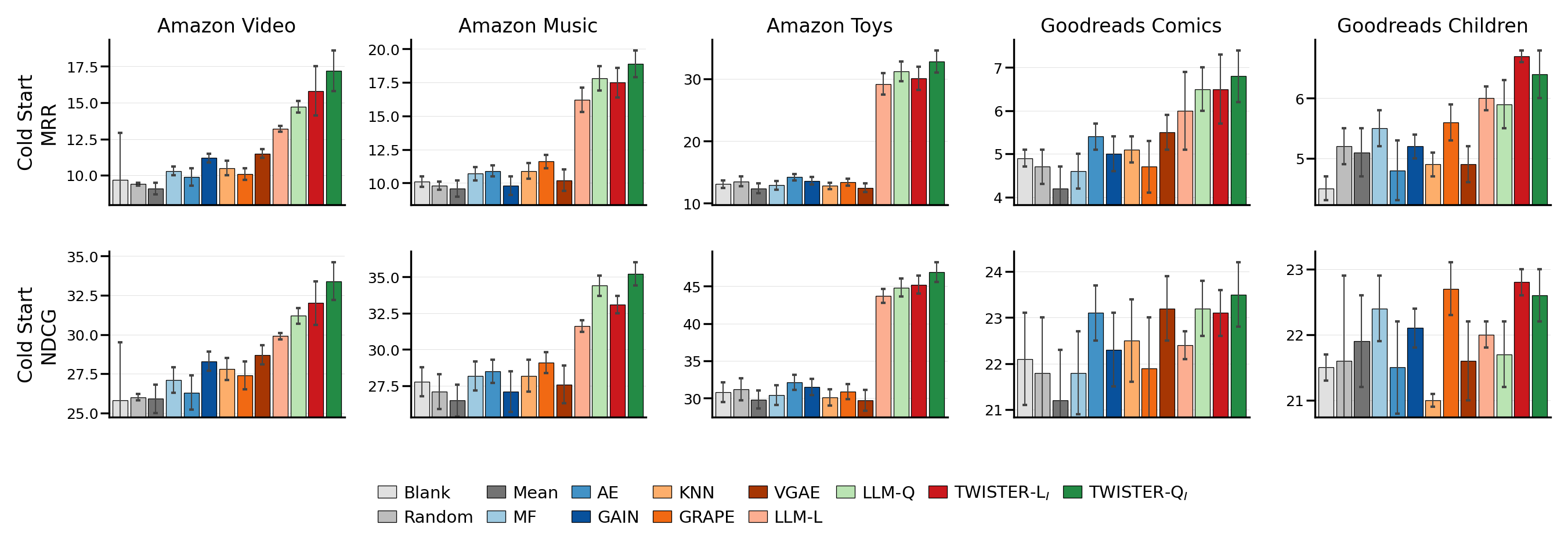}
    \vspace{-5ex}
    \caption{Recommendation performance under the \textit{Cold Start} setting ($50\%$ of users have no reviews in the training set).} 
    \label{fig:cold-start}
    \vskip -1ex
\end{figure*}
\subsubsection{Baselines.}
Follow prior works~\cite{zhang2024dataimputationperspectivegraph, you2020handlingmissingdatagraph}, we benchmark \method~ \\ against the following baselines (see Appendix \ref{ap_ssc:baseline} for details):
\vspace{-0.2em}
\begin{itemize}[leftmargin=1.5em]
\item \textbf{Blank / Mean / Random}: Standard numeric filling strategies;
\item \textbf{Structure-free Imputation}: Matrix Factorization (MF) \cite{kuleshov2015tensorfactorizationmatrixfactorization}, GAIN\cite{yoon2018gainmissingdataimputation}, and Autoencoder (AE)~\cite{bank2021autoencoders};
\item \textbf{Structure-based Imputation}: K-Nearest Neighbors (KNN), GRAPE~\cite{you2020handlingmissingdatagraph}, and Variational Graph Auto-Encoders (VGAE)~\cite{kipf2016variationalgraphautoencoders};
\end{itemize}

\noindent
We additionally compare with interaction-only
LightGCN~\cite{he2020lightgcn} (Supplementary Table~\ref{tab:lightgcn_twister_uniform}). We also evaluate several variants of our proposed approach in Section \ref{sec:ablation}, Ablation Studies:
\vspace{-0.2em}
\begin{itemize}[leftmargin=1.5em]

\item \textbf{\method{}}: the full method, with LLM-as-Graph-Aggregator on both user-side and item-side line graphs;
\item \textbf{\method$_U$}: user-side aggregation only (item-side removed);
\item \textbf{\method$_I$}: item-side aggregation only (user-side removed);
\item \textbf{\method$_{U+\text{meta}}$}: \method$_U$ augmented with structured item metadata in place of item-side aggregation;
\item \textbf{\texttt{LLM}}: generic LLM prompting with no graph structure (both views removed).

\end{itemize}
\vspace{-0.2em}
\noindent

Throughout the following subsections, we use \textbf{\method-L} and \textbf{\method-Q} to denote our method with \texttt{Llama3.2-3B} and \texttt{Qwen2.5-7B} backbones respectively~\cite{grattafiori2024llama3herdmodels-shorter,qwen2025qwen25technicalreport}. We defer the ablated variants (item-side-only, user-side-only, no-structure) to §\ref{sec:ablation}.

\subsection{Utility in Recommendation (RQ1)}
\label{ssec:rq1}

Tables~\ref{tab:rec_rr05} and Figure~\ref{fig:cold-start} report top-$k$
recommendation performance under the \emph{Uniform Masking} and
\emph{Cold-Start} protocols, respectively. Across all five domains, \method{} achieves the strongest
results.

\paragraph{Uniform Masking.}
When 50\% of reviews are randomly removed, \method{} delivers 
2--8\, absolute percentage-point (pp) NDCG gain over the best-performing baseline on the \textsc{Amazon} domains and up to 5\,pp on
\textsc{Goodreads}. Comparable improvements in ACC and AUC indicate that the
imputed reviews produced by \method{} not only read plausibly but also inject
useful ranking signals.

\paragraph{Cold Start.}
With 50\% of users lacking any reviews, user-side line graphs cannot
be constructed, so \method{} reduces to its item-side variant
\method$_I$ by construction (Supplementary Table~\ref{tab:rec_cold_start}).

Even in this harsher scenario, \method$_I$ surpasses the strongest graph baseline
(GRAPE) by 6\,pp NDCG on \textsc{Amazon} and 2\,pp on
\textsc{Goodreads}. This highlights the effectiveness of line-graph
aggregation for propagating contextual information to cold-start users.

\noindent\textbf{Takeaway.} \method{} consistently improves recommendation
quality under both random and cold-start setting, providing an affirmative
answer to \textbf{RQ1}.
\begin{figure}[t!]
    \centering
    \includegraphics[width=1\linewidth]{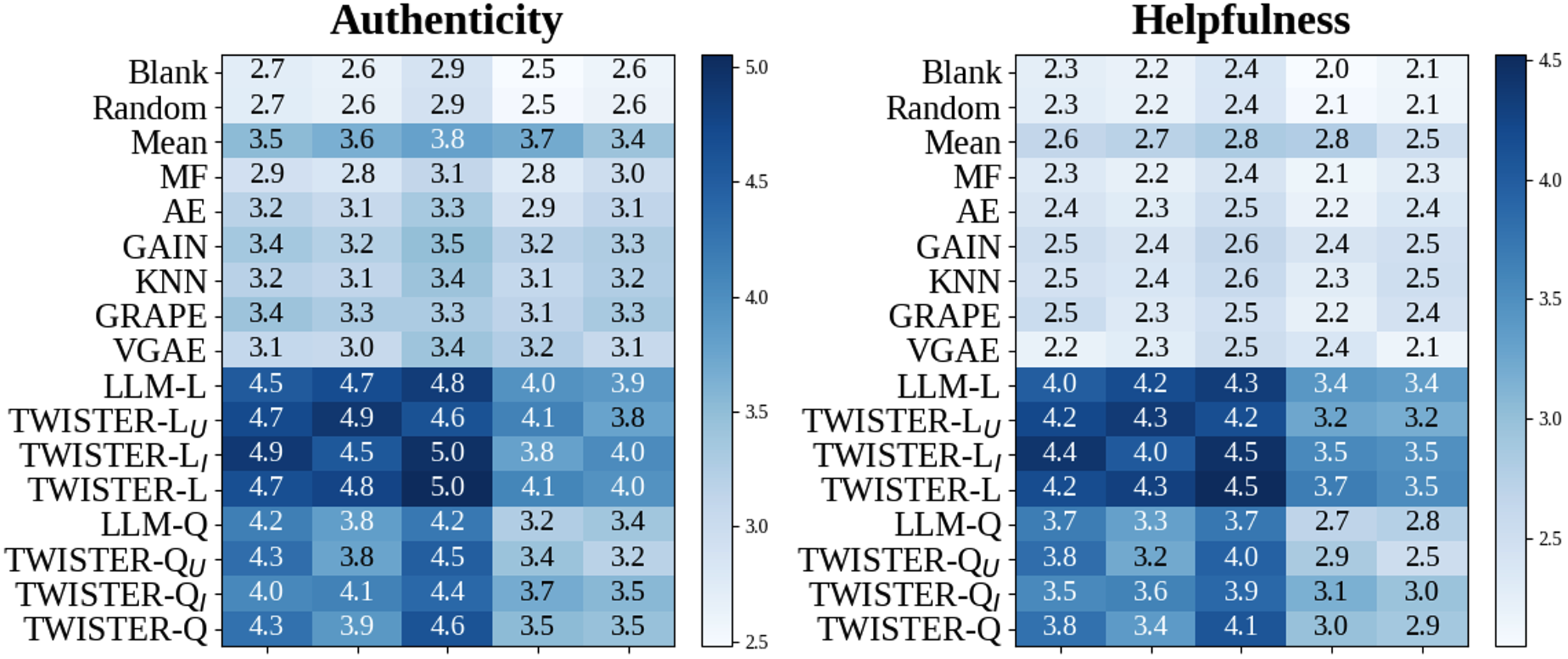}
    \caption{LLM-as-Judge scores (1--5) from DeepSeek-R1 on Authenticity and Helpfulness. Structure-aware variants (\method{}-L, \method{}-Q) consistently outperform. Full four-dimension results are in Appendix~\ref{sec:extra}.}

    \vskip -2ex
    \label{fig:llm-as-judge}
\end{figure}

\subsection{Semantic Awareness (RQ2)}
\label{ssec:rq2}

Here we probe whether \method{}'s imputed reviews are actually informative, not merely structurally conformant. We assess review-level quality with an LLM-as-Judge protocol (§\ref{sssec:llmjudge}) and similarity to the ground-truth reviews with ROUGE$_L$ and BERTScore (§\ref{ssec:semantic-fidelity}).

\subsubsection{LLM-as-Judge: Accessing Review Quality}
\label{sssec:llmjudge}

We follow recent practice by using large foundation models as automatic quality assessors. Specifically, we prompt \textsc{DeepSeek-R1} to rate each synthetic review on two key dimensions relevant for recommendation:

\begin{itemize}[leftmargin=1em]
    \item \textbf{Authenticity}: To resemble a genuine first-hand account.
    \item \textbf{Helpfulness}: To offer actionable advice for buyers.
\end{itemize}

Two additional dimensions (Specificity and Readability), the full
heatmap, and implementation details are provided in
Appendix~\ref{sec:extra}; the full evaluation prompt appears in
Table~\ref{tab:amazon_prompt}. We summarize our key findings in
Figure~\ref{fig:llm-as-judge} below:

\begin{enumerate}[leftmargin=1.5em]
\item \textbf{Structure-aware prompting 
improves quality.}
   Graph-based variants (\method$_U$, \method$_I$, the full \method{}) outperform all baselines, often by more than one point.
\item \textbf{Joint aggregation performs best.}
\method{} ranks top or near-top in all cases, leading on both authenticity and helpfulness ($\approx 4.4$) and showing the advantage of combining user and item context. Similar trends on two additional metrics---specificity and readability---appear in Appendix~\ref{sec:extra}.

\item \textbf{Traditional imputers lack depth.}
Methods like \texttt{Mean} and \texttt{KNN} produce fluent but shallow text, scoring low on authenticity ($\leq 3.5$)---shallow heuristics fail to capture product nuances even when surface fluency is adequate.

\end{enumerate}

\begin{figure}[t!]
    \centering
    \includegraphics[width=\linewidth]{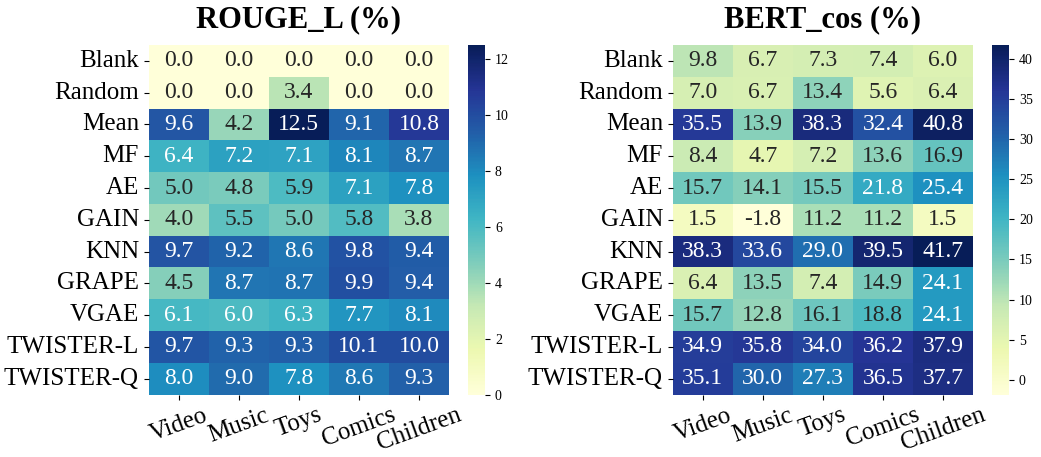}
    \vskip -0.5em
    \caption{Heatmap illustrating the semantic similarity of imputed and original texts. With the exception of the over-smoothed \texttt{KNN} and \texttt{Mean} baselines, our method (\texttt{\qw{\method-L}} and \texttt{\qw{\method-Q}}) achieves notably higher semantic fidelity.}
    \label{fig:fidelity}
    \vskip -2ex
\end{figure}

\begin{table*}[h]
\caption{Ablation studies under the \textit{Uniform Masking} setting.  The final column shows the mean rank of each method across the 20 dataset–metric cells (lower is better).}
\vskip -1em
\setlength{\tabcolsep}{1.1pt}
\renewcommand{\arraystretch}{1.15}
\centering
\begin{adjustbox}{width=1\textwidth}
\begin{tabular}{@{}lccccccccccccccccccccc@{}}
\toprule
        & \multicolumn{4}{c}{\textbf{Amazon\_Video}} & \multicolumn{4}{c}{\textbf{Amazon\_Music}} & \multicolumn{4}{c}{\textbf{Amazon\_Toys}} & \multicolumn{4}{c}{\textbf{Goodreads\_Comics}} & \multicolumn{4}{c}{\textbf{Goodreads\_Children}} & \textbf{Avg.}\\ 
\cmidrule(l){2-5} \cmidrule(l){6-9} \cmidrule(l){10-13} \cmidrule(l){14-17} \cmidrule(l){18-21} 
\textbf{Method } & ACC   & AUC   & MRR   & NDCG  & ACC   & AUC   & MRR   & NDCG  & ACC   & AUC   & MRR   & NDCG  & ACC    & AUC    & MRR    & NDCG   & ACC     & AUC     & MRR     & NDCG    & \textbf{Rank}\\ 
\midrule
\qw{\texttt{LLM-L}}   & 1.0$\pm$0.3 & 50.8$\pm$0.7 & 5.5$\pm$0.2 & 21.4$\pm$0.3 & 7.0$\pm$0.2 & 63.0$\pm$0.8 & 15.1$\pm$0.4 & 30.4$\pm$0.5 & 21.9$\pm$2.6 & 75.7$\pm$0.3 & 31.6$\pm$1.5 & 44.7$\pm$1.1 & 1.5$\pm$0.2 & 50.3$\pm$0.6 & 6.0$\pm$0.3 & 21.7$\pm$0.2 & 1.1$\pm$0.6 & 50.9$\pm$0.1 & 5.9$\pm$0.1 & 21.9$\pm$0.0 & 8.10 \\

\qw{\method-L$_I$} & 1.1$\pm$0.0 & 51.7$\pm$1.0 & 5.2$\pm$0.1 & 21.1$\pm$0.1 & \underline{8.8$\pm$0.3} & 68.0$\pm$0.9 & 17.1$\pm$0.1 & 32.4$\pm$0.3 & 22.4$\pm$1.9 & 73.7$\pm$0.3 & 32.0$\pm$1.4 & 44.8$\pm$1.0 & 1.6$\pm$0.2 & 51.5$\pm$0.5 & 6.1$\pm$0.8 & 21.8$\pm$0.3 & 1.5$\pm$0.5 & \underline{53.3$\pm$0.9} & \underline{6.7$\pm$0.2} & \underline{22.7$\pm$0.2} & 6.50 \\

\qw{\method-L$_U$} & \underline{1.7$\pm$0.7} & \underline{52.2$\pm$0.0} & \underline{8.6$\pm$2.6} & \underline{25.1$\pm$3.2} & \textbf{9.6$\pm$1.5} & \textbf{69.8$\pm$0.5} & \underline{18.4$\pm$1.7} & \underline{33.5$\pm$1.4} & 28.5$\pm$0.4 & \underline{83.3$\pm$0.1} & \underline{41.3$\pm$0.9} & \underline{53.2$\pm$0.8} & \underline{1.7$\pm$0.1} & \textbf{53.2$\pm$0.3} & \underline{6.2$\pm$0.1} & \underline{22.0$\pm$0.1} & \underline{2.1$\pm$0.7} & 51.3$\pm$0.3 & 6.6$\pm$0.6 & 22.5$\pm$0.5 & \underline{4.60} \\

\qw{\method-L$_{U+\text{meta}}$} & 1.0$\pm$0.0 & 51.5$\pm$0.1 & 5.8$\pm$0.0 & 21.7$\pm$1.7 & 8.4$\pm$1.8 & \underline{69.5$\pm$1.1} & \textbf{19.4$\pm$0.2} & \textbf{34.5$\pm$1.7} & \textbf{29.2$\pm$1.2} & \textbf{84.8$\pm$0.0} & \textbf{41.7$\pm$0.6} & \textbf{53.5$\pm$0.4} & 1.1$\pm$0.0 & 51.6$\pm$0.6 & 6.0$\pm$0.4 & 21.8$\pm$0.2 & 0.7$\pm$0.2 & \textbf{54.4$\pm$1.3} & 6.1$\pm$0.9 & 22.4$\pm$0.0 & 5.55 \\

\qw{\method-L} & \textbf{3.7$\pm$1.2} & \textbf{53.7$\pm$0.9} & \textbf{11.7$\pm$0.5} & \textbf{28.8$\pm$0.5} & 7.0$\pm$1.1 & 68.5$\pm$2.4 & 17.1$\pm$1.7 & 32.6$\pm$1.6 & \underline{28.9$\pm$3.5} & 80.7$\pm$0.8 & 40.2$\pm$2.8 & 51.9$\pm$2.2 & \textbf{1.8$\pm$0.3} & \underline{52.8$\pm$0.4} & \textbf{6.9$\pm$0.2} & \textbf{23.1$\pm$0.1} & \textbf{2.3$\pm$0.4} & 52.6$\pm$0.2 & \textbf{7.1$\pm$0.3} & \textbf{23.8$\pm$0.2} & \textbf{4.10}\\[2pt]
\midrule
\qw{\texttt{LLM-Q}}    & 2.1$\pm$1.0 & 51.9$\pm$0.8 & 5.7$\pm$0.3 & 21.5$\pm$0.2 & 5.6$\pm$0.3 & 62.7$\pm$0.4 & 14.6$\pm$0.2 & 29.9$\pm$0.3 & 5.8$\pm$1.9 & 58.4$\pm$1.5 & 18.7$\pm$1.8 & 35.6$\pm$1.5 & 2.4$\pm$0.1 & 52.9$\pm$0.2 & 6.4$\pm$0.0 & 22.4$\pm$0.0 & \textbf{4.0$\pm$0.2} & \textbf{54.5$\pm$0.3} & \textbf{7.8$\pm$0.2} & \textbf{25.6$\pm$0.3} & 6.25 \\

\qw{\method-Q$_I$}  & 1.7$\pm$0.6 & 51.7$\pm$0.4 & 8.7$\pm$0.3 & 25.2$\pm$0.2 & 5.3$\pm$0.2 & 62.9$\pm$0.0 & 15.1$\pm$0.1 & 29.7$\pm$0.4 & \underline{6.4$\pm$0.6} & \underline{60.1$\pm$1.7} & \underline{19.4$\pm$0.1} & 34.0$\pm$1.1 & 2.2$\pm$0.5 & 53.0$\pm$0.9 & 6.6$\pm$0.3 & 23.1$\pm$1.3 & 1.4$\pm$0.2 & 53.3$\pm$0.8 & 6.6$\pm$0.5 & 23.4$\pm$1.2 & 6.65 \\

\qw{\method-Q$_U$}  & \underline{4.9$\pm$2.4} & \underline{55.3$\pm$1.3} & \textbf{13.2$\pm$2.7} & \textbf{30.1$\pm$2.3} & 5.9$\pm$0.0 & 64.3$\pm$0.7 & 15.8$\pm$0.2 & 30.6$\pm$0.1 & 6.3$\pm$4.4 & 58.6$\pm$4.2 & 17.2$\pm$7.3 & 33.4$\pm$6.7 & 1.5$\pm$0.2 & 53.1$\pm$0.1 & 6.7$\pm$0.6 & 22.7$\pm$0.5 & 1.7$\pm$0.3 & \underline{54.3$\pm$0.3} & 7.2$\pm$0.3 & 23.9$\pm$0.7 & 5.45 \\[2pt]

\qw{\method-Q$_{U+\text{meta}}$} & 3.4$\pm$0.7 & 54.2$\pm$0.5 & 10.3$\pm$0.6 & 26.8$\pm$0.4 & \underline{8.9$\pm$0.9} & \underline{68.7$\pm$0.6} & \underline{18.8$\pm$0.3} & \textbf{34.2$\pm$0.5} & \textbf{7.8$\pm$1.4} & \textbf{62.1$\pm$0.7} & \textbf{21.4$\pm$1.1} & \textbf{36.2$\pm$0.8} & \underline{2.3$\pm$0.3} & \underline{53.8$\pm$0.2} & \underline{7.2$\pm$0.2} & \underline{24.0$\pm$0.3} & 1.7$\pm$0.4 & 53.1$\pm$0.3 & \underline{7.6$\pm$0.3} & \underline{25.4$\pm$0.2} & \underline{3.80} \\

\qw{\method-Q} & \textbf{5.9$\pm$0.5} & \textbf{59.7$\pm$1.1} & \underline{12.5$\pm$0.8} & \underline{27.9$\pm$0.7} & \textbf{9.6$\pm$0.1} & \textbf{69.6$\pm$0.2} & \textbf{19.0$\pm$0.3} & \underline{34.1$\pm$0.3} & 6.2$\pm$0.5 & 59.1$\pm$0.1 & 19.1$\pm$0.2 & \underline{34.9$\pm$0.8} & \textbf{2.6$\pm$0.4} & \textbf{54.2$\pm$0.3} & \textbf{7.3$\pm$0.2} & \textbf{24.1$\pm$0.2} & \underline{2.0$\pm$0.2} & 53.8$\pm$0.1 & 7.5$\pm$0.1 & 24.7$\pm$0.1 & \textbf{3.30} \\
\bottomrule
\end{tabular}
\end{adjustbox}
\label{tab:ablation}
\end{table*}

\noindent
Overall, Figure~\ref{fig:llm-as-judge} LLM-as-Judge results align with our metrics (§\ref{ssec:rq1}): \method{} reviews both boost performance and appear helpful and natural, highlighting the value of structure-aware generation.

\subsubsection{Semantic Fidelity}
\label{ssec:semantic-fidelity}
Figure~\ref{fig:fidelity} presents heatmaps comparing generated reviews with ground-truth using ROUGE\textsubscript{L} and BERT\textsubscript{cos}. Interestingly, although KNN and MEAN achieve higher surface-level similarity scores than \method, both exhibit inferior performance in assessment of LLM-as-Judge( $\S$\ref{sssec:llmjudge}) and downstream recommendation ($\S$\ref{ssec:rq1}), consistent with our earlier analysis that over-smoothness can degrade recommendation quality (see $\S$\ref{ssec:gap}). Aside from these baselines, \method\ consistently surpasses other embedding-based imputation methods across all datasets, indicating that jointly modeling structure and text yields richer lexical diversity and greater content coverage. Sentence-level BERTScore further corroborates these results, showing that reviews generated by \method\ are semantically closer to the ground-truth.

\subsection{Ablation Studies} \label{sec:ablation}

Table~\ref{tab:ablation} compares five structure-aware variants of our method using \texttt{Llama3.2-3B} and \texttt{Qwen2.5-7B} under a 50\% review-missing setting. The baseline \texttt{LLM} (no structure) performs worst across all metrics, confirming that structure-free generation is ineffective.

Incorporating line-graph context---either item-side (\method$_I$) or user-side (\method$_U$)---significantly improves performance. User-centric prompting (\method$_U$) consistently outperforms item-centric (\method$_I$), with average gains of 2.1 pp MRR and 2.3 pp NDCG on Llama, showing the importance of user-side aggregation.

Adding item metadata (\method$_{U+\text{meta}}$) yields further gains when such structured attributes are available. The best overall results come from combining both user and item contexts (\method{}), which consistently outperforms all other variants. In summary: (1) Graph context is crucial for effective imputation; (2) user-centric prompts are more informative than item-centric; and (3) joint views offer complementary gains and consistent improvements.

\section{TWISTER: Theoretical Foundations} 
\label{sec:theory}

Having established \method{}'s empirical effectiveness on recommendation utility (RQ1) and review quality (RQ2), we now develop a theoretical account of \emph{why} \method{}'s imputed reviews help downstream recommenders. We define structural smoothness via Dirichlet energy on the line graph (\S\ref{sec:dirichlet}) and relate it to the generalization error of a review-aware recommender (\S\ref{sec:bound}). Together, these tools predict that effective imputers should land in an \emph{intermediate}-smoothness regime---smooth enough for the bound in Proposition~\ref{prop:smooth-risk-new} to tighten, yet not so smooth as to collapse representation diversity. This motivates our third research question:
\begin{itemize}[leftmargin=2em,itemsep=0pt]
    \item[] \textbf{RQ3:} Does \method{}'s imputed review matrix exhibit the intermediate smoothness that our Dirichlet-energy analysis predicts for effective imputers?
\end{itemize}
We answer RQ3 empirically in \S\ref{sec:smoothness} and interpret the result via smoothness--variance trade-off in \S\ref{ssec:gap}. Notation here follow \S\ref{sec:method}.

\subsection{Dirichlet Energy on Line Graphs}
\label{sec:dirichlet}

Let $\mathbf{Z} = [\,\mathbf{z}_{e}\,]{e\in\Gamma}$ denote the review–embedding matrix, where each column $\mathbf{z}_{e}$ encodes the review $r_{e}$ for every observed user–item interaction $e$.
Let $\mathbf{L}_{U}$, $\mathbf{L}_{I}$, and $\mathbf{L}_{U,w}$ be the Laplacians of the \emph{user-side}, \emph{item-side}, and \emph{weighted user-side} line graphs, respectively, as defined in §\ref{sec:line-graphs}.
We measure smoothness for each view separately using \textit{Dirichlet Energy} (see Appendix \ref{sec:prelim-dirichlet} for details):

\begin{align}
\mathcal{E}_{U}(\mathbf Z)
      &=\operatorname{tr}\!\bigl(\mathbf Z^{\!\top}\mathbf L_{U}\mathbf Z\bigr)
        \;=\;\tfrac12\!
           \sum_{\substack{e_1\sim_U e_2}}\!
           \bigl\|\mathbf z_{e_1}-\mathbf z_{e_2}\bigr\|_2^2,
           \tag{U}\\[2pt]
\mathcal{E}_{U,w}(\mathbf Z)
      &=\operatorname{tr}\!\bigl(\mathbf Z^{\!\top}\mathbf L_{U,w}\mathbf Z\bigr)
        \;=\;\tfrac12\!
           \sum_{\substack{e_1\sim_{U} e_2}}\!
           W_{\ell_{e_1}\ell_{e_2}}
           \bigl\|\mathbf z_{e_1}-\mathbf z_{e_2}\bigr\|_2^2.
           \tag{U,w}
\end{align}

Here, $e_1 \sim_U e_2$ (respectively, $e_1 \sim_I e_2$) means that $e_1$ and $e_2$ are reviews written by the same user (respectively, reviews on the same item). $W_{\ell_{e_1}\ell_{e_2}}$ is a similarity weight between the items reviewed in $e_1$ and $e_2$, as defined in §\ref{sec:line-graphs}. The item-side energy $\mathcal{E}_I(\mathbf Z) = \operatorname{tr}(\mathbf Z^\top \mathbf L_I \mathbf Z)$ is defined analogously to $\mathcal{E}_U$.

\vspace{4pt}\noindent
\textbf{Interpretation.}
\begin{itemize}[leftmargin=*]
  \item \emph{Intra-user smoothness} ($\mathcal{E}_U$):  
        encourages a single user’s reviews to stay consistent across
        the different items they have rated.
  \item \emph{Inter-user consensus} ($\mathcal{E}_I$):  
        enforces coherence among reviews of the \emph{same item} to match crowd opinion.
  \item \emph{Fine-grained preference coherence} ($\mathcal{E}_{U,w}$):  
        further sharpens intra-user smoothness by weighting pairs of
        interactions—two reviews by the same user are required to be
        \emph{closer} when their items are semantically similar.
\end{itemize}

These three views decompose smoothness by neighbor type and will serve as diagnostics when we inspect $\widehat{\mathbf{R}}$ in §\ref{sec:smoothness}. The risk analysis that follows is stated in terms of a single, combined Dirichlet energy on the standard line graph, since that is the object $f_{\mathsf{T}}$ actually sees.

\subsection{Smoothness, Generalization, and its Pitfall}
\label{sec:bound}

Let $\Gamma$ be the set of observed user–item interactions (edges), and let
$e\in\Gamma$ index one such interaction with ground‑truth rating
$y_e\in\mathbb{R}$. 
Consider the linear recommender $f_{\mathbf w}:\mathbb{R}^{d}\!\to\!\mathbb{R}$,
\[
\hat y_e \;=\; f_{\mathbf w}(\mathbf z_e) \;=\; \mathbf z_e^{\!\top}\mathbf w,
\qquad \|\mathbf w\|_2 \le B,
\]
where $\mathbf w\in\mathbb{R}^{d}$ is the weight vector and $B>0$ bounds its
$\ell_2$ norm. We train with the squared loss and define the empirical risk as:
\[
\mathcal R(\mathbf w)
\;:=\;
\frac{1}{|\Gamma|}\sum_{e\in\Gamma}\ell(\hat y_e,y_e)
\;=\;
\frac{1}{2|\Gamma|}\sum_{e\in\Gamma}(\hat y_e - y_e)^2.
\]

Let $\mathbf{L} \in \mathbb{R}^{|\Gamma|\times|\Gamma|}$ be the Laplacian of the standard line graph $L(\mathcal{G})$ of the interaction graph, i.e., the graph whose nodes are the $|\Gamma|$ interactions and whose edges connect any two interactions sharing an endpoint (same user or same item). This is distinct from the specialized $\mathbf{L}_U$, $\mathbf{L}_I$, $\mathbf{L}_{U,w}$ used by the imputer in §\ref{sec:line-graphs}; $\mathbf{L}$ unions same-user and same-item relations into the single adjacency that mirrors what the downstream recommender $f_{\mathsf{T}}$ sees on $\mathcal{G}$
and has eigenvalues $0 = \lambda_1 \le \lambda_2 \le \cdots$, with $\lambda_{\min} := \min\{\lambda_k : \lambda_k > 0\}$. Then the (Dirichlet) smoothness of the review embeddings is
\[
 \mathcal{E}(\mathbf{Z})
\;:=\;
\operatorname{tr}\!\big(\mathbf{Z}^{\!\top}\mathbf{L} \mathbf{Z}\big)
\;=\;\tfrac12\sum_{e,e'\in\Gamma}A_{ee'}\|\mathbf z_e-\mathbf z_{e'}\|_2^2.
\]

We also write the empirical variance of ratings as
\[\operatorname{Var}(y):=\tfrac1{|\Gamma|}\sum_{e\in\Gamma}(y_e-\bar y)^2 \qquad \bar y:=\tfrac1{|\Gamma|}\sum_{e\in\Gamma} y_e \;.\]
\begin{proposition}[Smoothness controls prediction risk]
\label{prop:smooth-risk-new} 
Adopted from prior works \citep{Ando2007Laplacian,Zhou2011Laplacian,Chen2017BiasVariance},
let $\lambda_{\min}>0$ be the smallest non‑zero eigenvalue of \hspace{0.25ex}$\mathbf L$.
Then, for any $\mathbf w$ and review matrix $\mathbf Z$,
\vspace{-2pt}
\[
  \mathcal{R}(\mathbf{w})
  \;\le\;
  \frac{B^{2}}{|\Gamma|\,\lambda_{\min}}\,
  \mathcal{E}(\mathbf{Z})
  \;+\;
  \underbrace{\operatorname{Var}(y)}_{\text{irreducible noise}} \, .
\]
\end{proposition}

\noindent
\textit{Implication.}
Smaller Dirichlet energy tightens the bound, implying better expected
utility—\emph{provided the signal is not over‑smoothed.}

\vspace{6pt}
\noindent
\textbf{Over‑Smoothness Pitfall.}
Driving the energy toward zero is \emph{not} always beneficial:
\begin{itemize}[leftmargin=1.1em]
  \item \textbf{Constant fills} (e.g., Mean, Blank) erase user-specific nuance and degrade ranking accuracy.
  \item \textbf{KNN fills}
        suppress high‑frequency stylistic cues that encode fine‑grained
        preferences, likewise harming recommendations.
\end{itemize}

\noindent
Thus effective imputers must \emph{balance} smoothness and
expressivity.

\subsection{Structural Smoothness (RQ3)}
\label{sec:smoothness}

\begin{figure}[t!]
    \centering
     \includegraphics[width=0.85\linewidth]{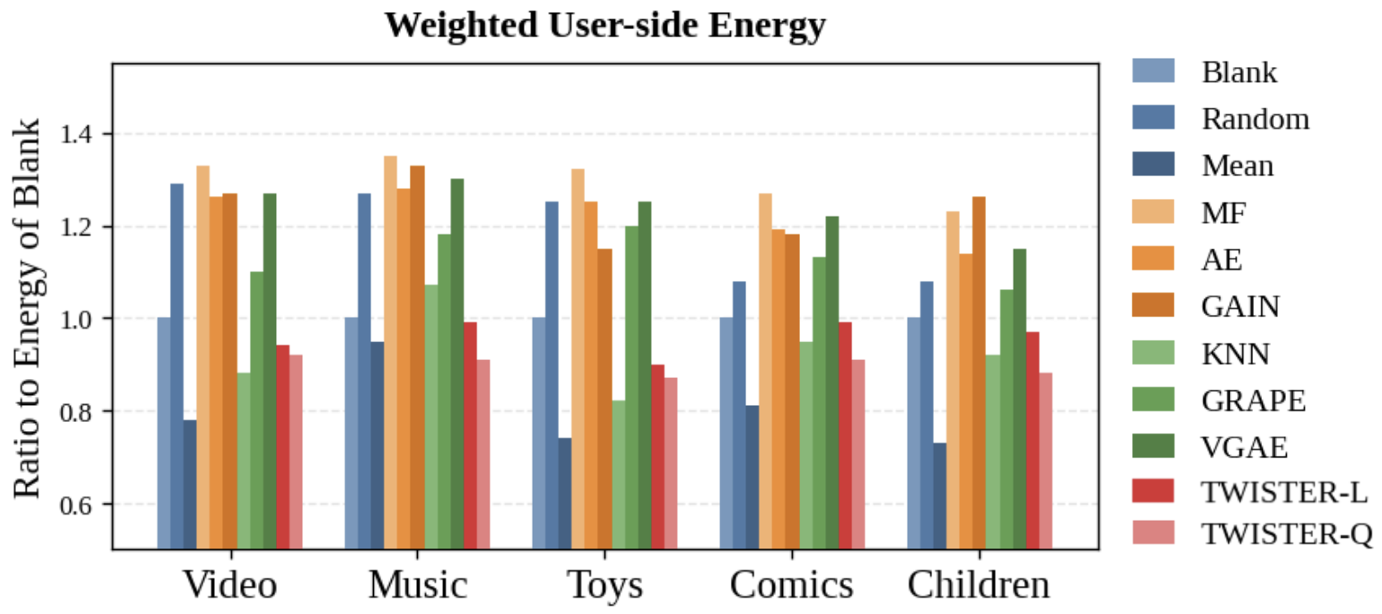}
  
      \vskip -1.0em
      \caption{Normalized Dirichlet energies on 
      weighted user-side line-graph views (lower is smoother). Excluding the over-smoothed \texttt{KNN} and \texttt{Mean} baselines, our methods \qw{(\method-L and \method-Q)} 
      attain the lowest energy.}
    \label{fig:energy}
    \vskip -2.5ex
\end{figure}

Figure~\ref{fig:energy} reports the weighted user-side Dirichlet energy $\mathcal{E}_{U,w}$, normalized to the \textbf{Blank} baseline so that lower values mean smoother signals; user- and item-side energies $\mathcal{E}_{U}$ and $\mathcal{E}_{I}$ appear in the Appendix~\ref{sec:extra} with the same pattern. Consistent with the semantic-fidelity analysis in §\ref{ssec:rq2}, \texttt{KNN} and \texttt{Mean} produce the lowest energies but deliver the weakest recommendation utility (§\ref{ssec:rq1})---the over-smoothing regime predicted by Proposition~\ref{prop:smooth-risk-new}.

Excluding \texttt{KNN} and \texttt{Mean}, \method{}-L and \method{}-Q attain the lowest energies. This indicates that \method{}'s imputed reviews adhere to the homophily structure the downstream recommender expects, while preserving the semantic fidelity that §\ref{ssec:rq2} showed the over-smooth baselines lack. Together, this confirms \textbf{RQ3}.

\subsection{Why \method{} Strikes the Right Balance}
\label{ssec:gap}

With the empirical results from §\ref{ssec:rq1}--§\ref{sec:smoothness} in hand, we now revisit the smoothness-variance trade-off introduced in §\ref{sec:bound} and show how \method{} positions itself within it. The theoretical prediction (Proposition~\ref{prop:smooth-risk-new}) and the experimental evidence converge on the same conclusion: effective imputation requires striking a balance, not maximizing smoothness.

\paragraph{Trade-off Principle.}
The generalization error $\mathcal{R}(\mathbf{w})$ of a linear recommender on $\mathbf{Z}$ is governed by the energy $\mathcal{E}(\mathbf{Z})$. However, pushing smoothness to the extreme ($\mathcal{E}(\mathbf{Z}) \to 0$) collapses representation diversity, limiting predictive power. Thus, there exists an optimal range for $\mathcal{E}(\mathbf{Z})$ that achieves a balance between smoothness and variation (see Proposition~\ref{prop:smooth-risk-new}).

\paragraph{Comparison of Imputation Strategies.}
We illustrate how different imputation approaches manage the smoothness–variance trade-off:

\begin{enumerate}[leftmargin=1.5em,label=(\alph*)]
  \item \textbf{Constant fills}: Yield large bias, resulting in representations that are overly smooth but inaccurate.
  \item \textbf{Random fills}: Are unbiased but suffer from high variance.
  \item \textbf{Structure-free models}: Ignore relational information, leading to persistently high variance.
  \item \textbf{Graph-based encoders}: Reduce variance but with semantic loss during text encoding.
  \item \textbf{\method{} (ours)}: Prompts an LLM with both user/item-side 
  contexts on the line graph. Here, prediction errors $X$ are assumed sub-Gaussian with proxy variance $\sigma_{\textsc{llm}}^2 \ll \sigma^2$, where $\sigma_{\textsc{llm}}^2$ ($\sigma^2$) denotes the error variance of \method{} (baseline):
     \vspace{-1ex}
    \[
      \Pr(|\delta| \geq t) \leq 2\exp\left(-\frac{t^2}{2\sigma_{\textsc{llm}}^2}\right)
      \quad\text{for all } t > 0.
    \]

    \vspace{-0.5ex}
   Here, $t$ is a positive threshold for the prediction error $|\delta|$. As a result, \method{} attains a balanced energy that preserves structural signals, thus avoiding the detrimental collapse observed for Mean and \textsc{KNN} in Proposition~\ref{prop:smooth-risk-new}.
\end{enumerate}

In summary, \method{} strikes an effective balance between smoothness and variance, leading to improved generalization performance (i.e., lower $\mathcal{R}(\mathbf{w})$).

\noindent
This explains the empirical observation in §\ref{sec:exp} (full details and proofs are in Appendix~\ref{app_sec:add_theory}):
\begin{itemize}[leftmargin=*]
    \item \textbf{Over-smoothed methods} (e.g., Mean, KNN) yield lower $\mathcal{E}(\mathbf{Z})$ and higher similarity with original text, but underperform in recommendation due to loss of expressivity.
    \item \textbf{Other baselines} either lack structural alignment (high $\mathcal{E}(\mathbf{Z})$) or underutilize graph information.
    \item   \textbf{\method{}} achieves lowest recommendation error by placing $\mathcal{E}(\mathbf{Z})$ in the “Goldilocks zone,” balancing graph-based smoothness with semantic richness.
\end{itemize}

\section{Related Works}\label{sec:related}

Having established the empirical utility of our framework  
for recommendation under sparse review text, we next position it relative to prior work in data imputation and review-aware recommendation.
\paragraph{Feature Imputation}
Early work frames imputation as a \textit{matrix completion} problem, leveraging low-rank assumptions to recover missing entries~\cite{candes2012exact,hastie2015matrix, zhao2021action}. Subsequent \textit{deep learning }approaches employ autoencoders~\cite{costa2018missing,abiri2019establishing,gondara2018mida,vincent2008extracting} to capture nonlinear feature correlations, but are generally limited to dense numeric or categorical data and overlook structural information. To encode relational inductive biases, \textit{graph-based} methods such as GRAPE~\cite{you2020handlingmissingdatagraph} formulate imputation as edge prediction in bipartite graphs, while extensions such as variational graph autoencoders~\cite{kipf2016variationalgraphautoencoders} further enhance expressiveness. 
However, these methods either lose semantic richness by embedding texts as dense vectors or overlook structural dependencies. In contrast, our setting requires preserving both edge-level language and interaction structure, rather than imputing dense numeric features alone.

\vspace{-1.5ex}
\paragraph{Textual-Edge Graphs (TEGs) in Recommendation}
TEGs explicitly associate natural language content with edges, supporting joint reasoning over semantics and graph structure. The first formal benchmark for TEGs was introduced by~\citet{li2024teg}. Review-aware recommendation models operating on TEGs, such as Edgeformers~\cite{jin2023edgeformers} and Link2Doc~\cite{ling2024linkpredictiontextualedge}, demonstrate improvements in downstream performance by incorporating review text into graph-based architectures. However, \emph{most existing methods assume every user–item interaction has a textual review}, which is unrealistic—real-world review data is highly sparse.
This leaves open the problem of recommendation under sparse review text, where edge-level textual signals are only partially observed. Our work addresses this setting through structure-aware review synthesis on TEGs.

\section{Conclusion}
\label{sec:conclusion}
In this work, we present \method{}, a unified framework for imputing missing reviews for recommendation under sparse review text. By representing user--item interactions as a Textual-Edge Graph and deploying an LLM as a graph-aware aggregator over user- and item-side line-graph views, TWISTER produces reviews that are both grounded in the specific interaction and informed by relational context. 
Experiments across five Amazon and Goodreads benchmarks show consistent gains in recommendation utility and review quality over strong numeric, graph-based, and LLM baselines; a Dirichlet-energy analysis further shows that TWISTER's completions occupy the intermediate-smoothness regime our theory identifies as beneficial for downstream recommenders.

\textbf{Future directions.} Two natural extensions are (i) learned neighbor selection that adaptively weighs which interactions contribute to each $\phi^\star$, and (ii) joint training of the imputer and recommender with an explicit Dirichlet-energy regularizer.

\appendix



\begin{figure}[t!]
    \centering
    \includegraphics[width=1\linewidth]{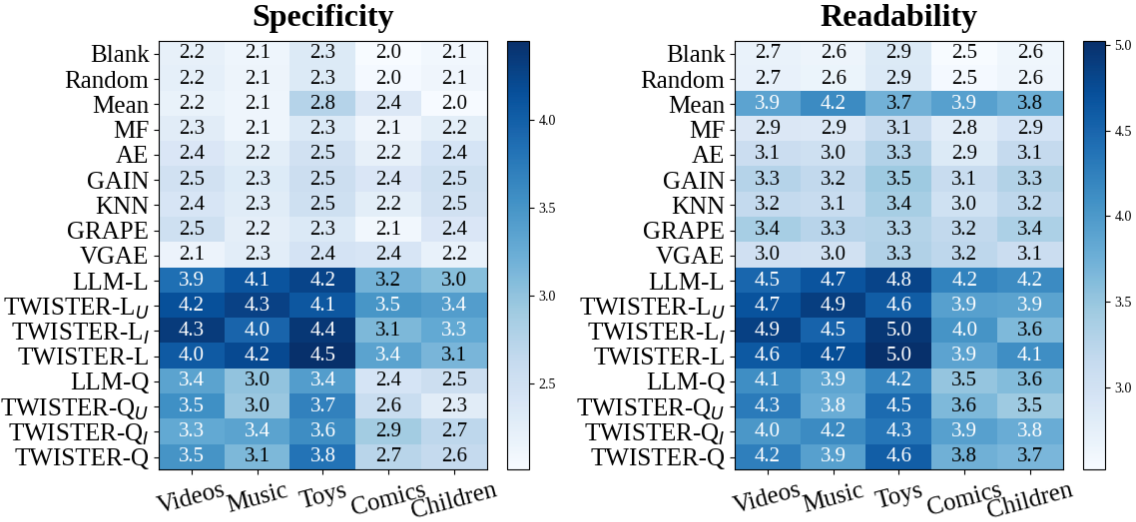}
     \vskip -1em
    \caption{
\textbf{LLM-as-Judge heatmaps.}
DeepSeek-R1 scores (1--5) for synthetic reviews on Specificity and Readability across methods/datasets. 
Structure-aware LLM variants (\texttt{\qw{\method-L}}, \texttt{\method-Q}) consistently outperform the baselines.
}

    \vskip -1.2em
    \label{fig:llm-as-judge_4}
\end{figure}

\section{Implementation Details}
\label{app_subsec:code}


\subsection{Models and Hyperparameters}
\label{app_ssec:model}
We utilize \texttt{Llama3.2-3B-Instruct} \cite{grattafiori2024llama3herdmodels-shorter} and \texttt{Qwen2.5-7B-Instruct} \cite{qwen2025qwen25technicalreport} as our text generation models. For embedding-based methods, we use Sentence-BERT (\texttt{gtr-t5-base}) as the text encoder and apply vec2text~\cite{morris-etal-2023-text} to decode embeddings back into text. For downstream recommendation tasks, Edgeformers~\cite{jin2023edgeformers} serves as our backbone model. For two text generation model, we set max new token as 256. 


\subsection {Computing Environment and Resources}
We implement \method~ under PyG \cite{fey2019fastgraphrepresentationlearning} and Sentence Transformer \cite{reimers2019sentence} modules.
Experiments are conducted on a NVIDIA GeForce RTX 4090 and the OS was Ubuntu 22.04.4 LTS with 128GB RAM.

\subsection{Datasets} Below, we present the dataset abbreviations together with details of the data preprocessing procedure.
\begin{itemize}[leftmargin=1em]
 \item \textit{Amazon\_Video\_Games} (shortened as \texttt{Amazon\_Video} or \texttt{Video}),  
 \item \textit{Amazon\_Musical\_Instruments} (\texttt{Amazon\_Music} or \texttt{Music}), 
  \item \textit{Amazon\_Toys\_And\_Games} (\texttt{Amazon\_Toys} or \texttt{Toys}), 
   \item \textit{Goodreads\_Comics \& Graphic} (\texttt{Comics}), and 
    \item \textit{Goodreads\_Children} (\texttt{Children}).
\end{itemize}

\textbf{Data Preprocessing.}
Each dataset is first converted into a bipartite TEG as described in $\S$~\ref{sec:bipartite-teg}. We then extract the $k$-core, retaining only those vertices involved in at least $k$ interactions. For efficiency, we further sample an \emph{ego} subgraph by selecting $100$ seed users uniformly at random and including their one-hop neighbors, resulting in a compact yet representative subset. We present the dataset statistics in Table~\ref{tab:dataset}.

\vspace{-1ex}
\section{Baseline Details}
\label{ap_ssc:baseline}

\paragraph{Numeric filling.}
\textbf{Blank} leaves missing reviews empty,
\textbf{Mean} fills them using the observed feature-wise mean, and
\textbf{Random} uses randomly generated strings of matched lengths.

\vspace{-1ex}
\paragraph{Structure-free imputation.}
\textbf{MF}~\cite{kuleshov2015tensorfactorizationmatrixfactorization}
uses low-rank matrix reconstruction;
\textbf{GAIN}~\cite{yoon2018gainmissingdataimputation} uses adversarial
imputation; and
\textbf{AE}~\cite{bank2021autoencoders} reconstructs missing values
through an autoencoder.

\vspace{-1ex}
\paragraph{Structure-based imputation.}
\textbf{KNN} imputes from nearest observed samples,
\textbf{GRAPE}~\cite{you2020handlingmissingdatagraph} exploits graph
structure for imputation, and
\textbf{VGAE}~\cite{kipf2016variationalgraphautoencoders} models
structural dependencies through a variational graph autoencoder.

\vspace{-1ex}
\section{Additional Experimental Results}
\label{sec:extra}

\begin{table}[t]
\caption{Statistics of the raw data. }
\footnotesize  
\vskip -1.5em
\begin{tabular}{@{}ccll} 
\toprule
Datasets & \# Reviews& \# Item&\# User\\ \midrule
Videos   &          2,565,3629& 71,982&1540,618\\
Music    &          1,512,530& 11,2222&90,330\\
Toys     &          8,194,101& 624,792&4,204,994\\
Comics   &          542,015& 89,311&59,347\\
Children &          734,640& 123,946&92,667\\ \bottomrule
\end{tabular}
\vskip -0.75em
\label{tab:dataset}
\end{table}

We prompt \textsc{DeepSeek-R1} to rate each synthetic review on four key dimensions relevant for recommendation: \textbf{Authenticity}: resembles a genuine first-hand account; \textbf{Helpfulness}: offers actionable advice for buyers; \textbf{Specificity}: provides concrete product details, not just general sentiment; and \textbf{Readability}: reads clear, coherent, and easy to follow.
Scores range from 1--5 (1--2 low, 3 acceptable, 4--5 high). We use three random seeds and report the mean per (method, dataset, dimension); standard deviations were always below 0.05 and are therefore omitted.

Figure~\ref{fig:llm-as-judge_4} extends Figure~\ref{fig:llm-as-judge} with the additional Specificity and Readability dimensions. The same qualitative patterns hold: structure-aware \method{} variants lead across all four dimensions, while shallow baselines (\texttt{Mean}, \texttt{KNN}) produce fluent but shallow text with low authenticity and specificity scores. The complete LLM-as-Judge evaluation prompt is provided in
Supplementary Section~\ref{sec:llm_judge_prompt}.

\section*{Ethical Considerations}

Our work studies synthetic review imputation for review-aware recommendation using public Amazon and Goodreads benchmarks. While \method{} is intended to recover missing textual signals for downstream recommendation, its generation capability could be misused to create deceptive reviews or manipulate recommender systems. Synthetic reviews should therefore remain internal to the recommendation pipeline or be clearly identified as machine-generated rather than presented as user-authored content.

Because \method{} aggregates neighboring user and item reviews, it may reproduce biases, inappropriate content, or sensitive information from source data. Inaccurate generations may also introduce spurious preference signals and affect downstream rankings. Practical deployments should therefore incorporate content filtering, privacy-preserving data handling, access controls, and auditing of recommendation outcomes. Our experiments use only existing public benchmark datasets and involve no new collection of user data.
\bibliographystyle{ACM-Reference-Format}
\bibliography{reference}
\balance 

\clearpage

\section*{Supplementary Material}
\setcounter{section}{0}
\renewcommand{\thesection}{S\arabic{section}}
\renewcommand{\thesubsection}{\thesection.\arabic{subsection}}
\renewcommand{\theHsection}{S\arabic{section}}
\renewcommand{\theHsubsection}{S\arabic{section}.\arabic{subsection}}

\section{Efficiency Analysis}
In this section, we report the LLM generation time. The time of LLM-as-Graph-Aggregrator ($\S$ \ref{sec:llm-conv-full}) is included in Table \ref{tab:separated_summary_analysis}, and the time for LLM-based imputation ($\S$ \ref{sec:edge-impute}) is included in Table \ref{tab:imputation_analysis}.
\begin{table}[h] 
    \centering
    \footnotesize   
        \setlength{\tabcolsep}{0.75pt}
       \renewcommand{\arraystretch}{0.5}
    \caption{Aggregation Token Counts and Generation Time}
    \vskip -0.5em
    \label{tab:separated_summary_analysis}
    \begin{tabular}{llcrrrrr}
    \toprule
    \multirow{2}{*}{\textbf{Dataset}} & \multirow{2}{*}{\textbf{Model}} & \multirow{2}{*}{\textbf{Type}} & \multirow{2}{*}{\textbf{Count}} & \multicolumn{2}{c}{\textbf{Total}} & \multicolumn{2}{c}{\textbf{Average}} \\
    \cmidrule(lr){5-6} \cmidrule(lr){7-8}
    & & & & \textbf{Tokens} & \textbf{Time (s)} & \textbf{Tokens} & \textbf{Time (s)} \\
    \midrule
\multirow{4}{*}{Children} & \multirow{2}{*}{Llama} & User & 70 & 11,889 & 59.4 & 169.8 & 0.849 \\
 & & Item & 1,326 & 115,004 & 575.0 & 86.7 & 0.434 \\
 & \multirow{2}{*}{Qwen} & User & 70 & 9,273 & 46.4 & 132.5 & 0.662 \\
 & & Item & 1,326 & 94,303 & 471.5 & 71.1 & 0.356 \\
     \midrule
\multirow{4}{*}{Comics} & \multirow{2}{*}{Llama} & User & 70 & 12,118 & 60.6 & 173.1 & 0.866 \\
 & & Item & 956 & 82,802 & 414.0 & 86.6 & 0.433 \\
 & \multirow{2}{*}{Qwen} & User & 70 & 9,088 & 45.4 & 129.8 & 0.649 \\
 & & Item & 956 & 70,381 & 351.9 & 73.6 & 0.368 \\
     \midrule
\multirow{4}{*}{Game} & \multirow{2}{*}{Llama} & User & 100 & 14,774 & 73.9 & 147.7 & 0.739 \\
 & & Item & 1,044 & 71,175 & 355.9 & 68.2 & 0.341 \\
 & \multirow{2}{*}{Qwen} & User & 100 & 12,262 & 61.3 & 122.6 & 0.613 \\
 & & Item & 1,044 & 56,228 & 281.1 & 53.9 & 0.269 \\
     \midrule
\multirow{4}{*}{Music} & \multirow{2}{*}{Llama} & User & 99 & 12,095 & 60.5 & 122.2 & 0.611 \\
 & & Item & 330 & 23,261 & 116.3 & 70.5 & 0.352 \\
 & \multirow{2}{*}{Qwen} & User & 99 & 9,313 & 46.6 & 94.1 & 0.470 \\
 & & Item & 330 & 18,841 & 94.2 & 57.1 & 0.285 \\
     \midrule
\multirow{4}{*}{Toys} & \multirow{2}{*}{Llama} & User & 100 & 12,486 & 62.4 & 124.9 & 0.624 \\
 & & Item & 1,002 & 51,770 & 258.9 & 51.7 & 0.258 \\
 & \multirow{2}{*}{Qwen} & User & 100 & 9,988 & 49.9 & 99.9 & 0.499 \\
 & & Item & 1,002 & 39,345 & 196.7 & 39.3 & 0.196 \\
    \bottomrule
    \end{tabular}%
    \vspace{-2.5ex}
\end{table}

\begin{table}[h!]
    \centering
   
    \setlength{\tabcolsep}{1.5pt}
    \caption{Imputation Generation Token Counts and Time}
    \vskip -0.5em
    \label{tab:imputation_analysis}
    \resizebox{0.95\columnwidth}{!}{%
    \begin{tabular}{llrrrrr}
    \toprule
    \multirow{2}{*}{\textbf{Dataset}} & \multirow{2}{*}{\textbf{Model}} & \multirow{2}{*}{\textbf{\# Review }}& \multicolumn{2}{c}{\textbf{Total Metrics}} & \multicolumn{2}{c}{\textbf{Average Metrics}} \\
    \cmidrule(lr){4-5} \cmidrule(lr){6-7}
    & & & \textbf{Imputation} & \textbf{Imputation} & \textbf{Imputation} & \textbf{Imputation} \\
    & & & \textbf{Tokens} & \textbf{Time (s)} & \textbf{Tokens} & \textbf{Time (s)} \\
    \midrule
    \multirow{2}{*}{Children} & Llama & 3,714 & 272,902 & 1,364.5 & 73.5 & 0.367 \\
     & Qwen & 3,714 & 260,770 & 1,043.1 & 70.2 & 0.281 \\
     \midrule 
    \multirow{2}{*}{Comics} & Llama & 2,310 & 327,866 & 1,639.3 & 141.9 & 0.710 \\
     & Qwen & 2,310 & 188,131 & 752.5 & 81.4 & 0.326 \\
       \midrule 
    \multirow{2}{*}{Games} & Llama & 1,334 & 297,443 & 1,487.2 & 223.0 & 1.115 \\
     & Qwen & 1,334 & 221,571 & 886.3 & 166.1 & 0.664 \\
       \midrule 
    \multirow{2}{*}{Music} & Llama & 926 & 105,127 & 525.6 & 113.5 & 0.568 \\
     & Qwen & 926 & 58,429 & 233.7 & 63.1 & 0.252 \\
       \midrule 
    \multirow{2}{*}{Toys} & Llama & 1,132 & 128,856 & 644.3 & 113.8 & 0.569 \\
     & Qwen & 1,132 & 77,463 & 309.9 & 68.4 & 0.274 \\
    \bottomrule
    \end{tabular}%
    }
    \vspace{-1.5ex}
\end{table}




\section{Additional Theory and Proofs}
\label{app_sec:add_theory}

\subsection{ Dirichlet Energy and Structural Smoothness}
\label{sec:prelim-dirichlet}
We briefly review the Dirichlet energy on graphs, a standard measure of signal smoothness in spectral graph theory~\cite{smola2003kernels, chung1997spectral,zhou2005regularization}.

Let $\mathcal H=(\mathcal V,\mathcal E,W)$ be any undirected graph with
non-negative edge weights $W:\mathcal E\!\to\!\mathbb R_{\ge0}$ and
$\mathbf L:=\mathbf D-\mathbf W$ be its unnormalised Laplacian
($\mathbf D$ diagonal, $D_{vv}=\sum_{v'}W_{vv'}$,
$\mathbf W$ the weighted adjacency). 

\vspace{3pt}\noindent
\textbf{Node signal.}
For a matrix
$\mathbf X=[\,\mathbf x_v\,]_{v\in\mathcal V}\in\mathbb R^{|\mathcal V|\times d}$
whose rows encode a $d$-dimensional feature vector per node, the
\emph{Dirichlet energy} of $\mathbf X$ on $\mathcal H$ is
\begin{equation}
E_{\mathcal H}(\mathbf X)
   :=\operatorname{tr}\!\bigl(\mathbf X^{\!\top}\mathbf L\,\mathbf X\bigr)
   =\tfrac12\!\sum_{\{v,v'\}\in\mathcal E}
      W_{vv'}\,\bigl\|\mathbf x_v-\mathbf x_{v'}\bigr\|_2^2.
\label{eq:dirichlet-prelim}
\end{equation}

\vspace{1pt}\noindent
\textbf{Smoothness interpretation.}
$E_{\mathcal H}(\mathbf X)$ quantifies how rapidly the signal
$\mathbf X$ varies across adjacent nodes:
\begin{itemize}[leftmargin=*]
  \item $E_{\mathcal H}(\mathbf X)=0$  
        $\;\Longleftrightarrow\;$  
        $\mathbf x_v=\mathbf x_{v'}$ for every edge $\{v,v'\}$  
        $\;\Longleftrightarrow\;$  
        $\mathbf X$ is \emph{piece-wise constant} on each connected
        component of $\mathcal H$.
  \item Smaller energy $\Longrightarrow$ higher \emph{structural
        smoothness}, i.e.\ neighbouring nodes carry more similar
        features.
\end{itemize}

\begin{figure}[H]
    \centering
    \includegraphics[width=0.8\linewidth]{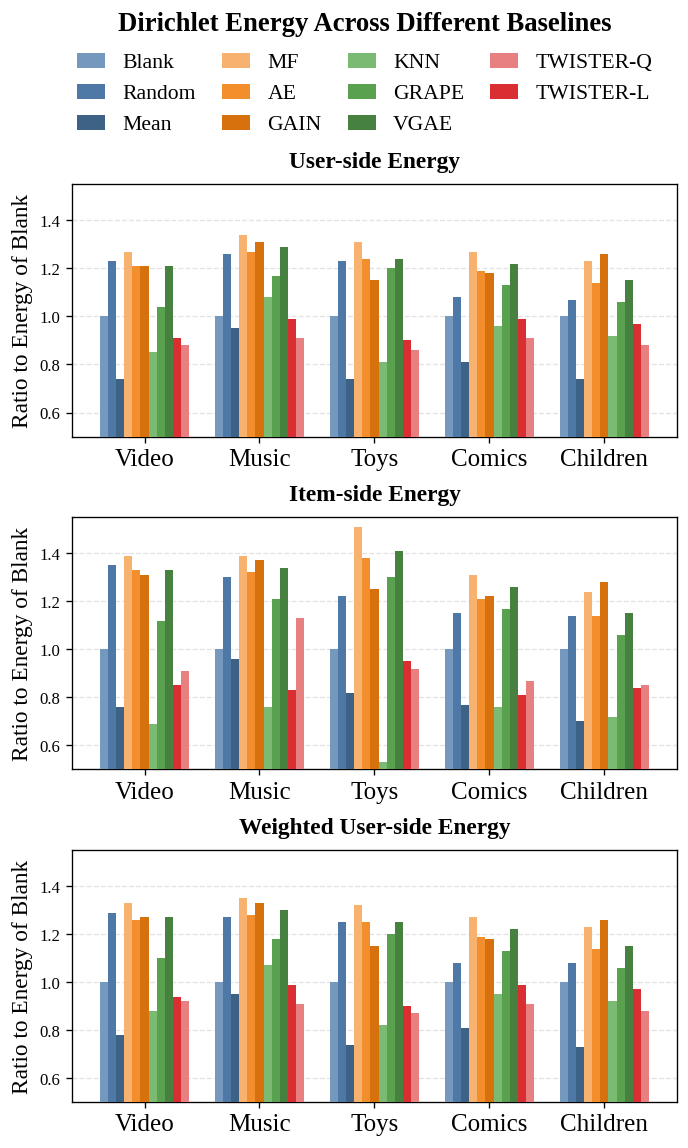}
    \vspace{-1ex}
      \caption{Normalized Dirichlet energies on
      user-, item-, and weighted user-side line-graph views (lower is smoother). Excluding the over-smoothed \texttt{KNN} and \texttt{Mean} baselines, our methods \qw{(\method-L and \method-Q)} 
      attain the lowest energy.}
    \label{fig:energy_3}
    \vskip -1ex
\end{figure}

\subsection{Risk Bound and Smoothness}
Let $\mathbf{L}$ be the Laplacian, and $B$ the norm bound for $\mathbf{w}$.
\begin{proposition}[Smoothness controls prediction risk]
For any $\mathbf{w}$ with $\|\mathbf{w}\|_2 \le B$,
\[
\mathcal{R}(\mathbf{w})
\leq \frac{B^2}{|\Gamma| \lambda_{\min}}\,E(\mathbf{Z}) + \operatorname{Var}(y)
\]
where $\lambda_{\min}$ is the smallest nonzero eigenvalue of $\mathbf{L}$.
\end{proposition}

\paragraph{Proof sketch.} This follows by relating the variance of predictions $\hat{y}_e = \mathbf{z}_e^\top \mathbf{w}$ to the energy $E(\mathbf{Z})$ via the Poincaré inequality, then using Jensen's inequality and standard bias-variance decomposition.

\clearpage

\begin{table*}[t]
\setlength{\tabcolsep}{1.2pt}
\centering
\caption{\qw{Recommendation performance under the \textit{Cold Start} setting (50\% of users have no reviews in the training set). In addition to the MRR and NDCG metrics shown in Figure~\ref{fig:cold-start}, we report the full four metrics (ACC, AUC, MRR, NDCG) here.}}
\vskip -1em
\begin{adjustbox}{width=1\textwidth}
\begin{tabular}{@{}lccccccccccccccccccccc@{}}
\toprule
& \multicolumn{4}{c}{\textbf{Amazon\_Video}} & \multicolumn{4}{c}{\textbf{Amazon\_Music}} & \multicolumn{4}{c}{\textbf{Amazon\_Toys}} & \multicolumn{4}{c}{\textbf{Goodreads\_Comics}} & \multicolumn{4}{c}{\textbf{Goodreads\_Children}} & \textbf{Avg.}\\ 
\cmidrule(l){2-5} \cmidrule(l){6-9} \cmidrule(l){10-13} \cmidrule(l){14-17} \cmidrule(l){18-21}
\textbf{Method} & ACC & AUC & MRR & NDCG & ACC & AUC & MRR & NDCG & ACC & AUC & MRR & NDCG & ACC & AUC & MRR & NDCG & ACC & AUC & MRR & NDCG & \textbf{Rank}\\
\midrule
Blank   & 3.5$\pm$1.4 & 51.8$\pm$0.7 &  9.7$\pm$3.2 & 25.8$\pm$3.7 & 3.5$\pm$0.3 & 51.6$\pm$1.7 & 10.1$\pm$0.4 & 27.8$\pm$1.0 & 4.6$\pm$1.1 & 54.1$\pm$1.2 & 13.1$\pm$0.6 & 30.8$\pm$1.3 & 0.9$\pm$0.2 & 51.2$\pm$1.4 & 4.9$\pm$0.2 & 22.1$\pm$1.0 & 0.8$\pm$0.3 & 51.0$\pm$0.1 & 4.5$\pm$0.2 & 21.5$\pm$0.2 & 10.00 \\
Random  & 3.2$\pm$0.5 & 50.9$\pm$1.2 &  9.4$\pm$0.1 & 26.0$\pm$0.2 & 3.1$\pm$0.4 & 51.2$\pm$1.5 &  9.8$\pm$0.3 & 27.1$\pm$1.2 & 4.8$\pm$1.2 & 54.5$\pm$1.4 & 13.5$\pm$0.8 & 31.2$\pm$1.5 & 0.7$\pm$0.3 & 50.9$\pm$1.6 & 4.7$\pm$0.4 & 21.8$\pm$1.2 & 1.0$\pm$0.4 & 51.5$\pm$0.9 & 5.2$\pm$0.3 & 21.6$\pm$1.3 &  9.93 \\
Mean    & 3.4$\pm$0.7 & 50.8$\pm$1.2 &  9.1$\pm$0.4 & 25.9$\pm$0.9 & 2.9$\pm$0.5 & 50.7$\pm$1.3 &  9.6$\pm$0.6 & 26.5$\pm$1.1 & 4.1$\pm$0.9 & 53.2$\pm$1.4 & 12.4$\pm$0.8 & 29.8$\pm$1.2 & 0.6$\pm$0.4 & 50.6$\pm$1.7 & 4.2$\pm$0.5 & 21.2$\pm$1.1 & 1.2$\pm$0.3 & 51.8$\pm$0.9 & 5.1$\pm$0.4 & 21.9$\pm$0.7 & 11.47 \\
\midrule
MF      & 3.8$\pm$0.5 & 52.1$\pm$0.9 & 10.3$\pm$0.3 & 27.1$\pm$0.8 & 3.6$\pm$0.4 & 51.9$\pm$1.5 & 10.7$\pm$0.5 & 28.2$\pm$1.0 & 4.3$\pm$0.8 & 53.5$\pm$1.1 & 12.9$\pm$0.7 & 30.4$\pm$1.3 & 0.8$\pm$0.3 & 50.9$\pm$1.4 & 4.6$\pm$0.4 & 21.8$\pm$0.9 & \textbf{1.3$\pm$0.2} & 52.0$\pm$0.6 & 5.5$\pm$0.3 & 22.4$\pm$0.5 &  8.38 \\
AE      & 3.2$\pm$0.8 & 51.6$\pm$1.3 &  9.9$\pm$0.6 & 26.3$\pm$1.1 & 3.9$\pm$0.3 & 52.3$\pm$1.2 & 10.9$\pm$0.4 & 28.5$\pm$0.8 & 5.1$\pm$0.7 & 54.6$\pm$0.9 & 14.2$\pm$0.5 & 32.1$\pm$1.0 & 1.2$\pm$0.2 & \underline{51.7$\pm$1.1} & 5.4$\pm$0.3 & 23.1$\pm$0.6 & 0.9$\pm$0.4 & 50.8$\pm$0.8 & 4.8$\pm$0.5 & 21.5$\pm$0.7 &  7.75 \\
GAIN    & 4.1$\pm$0.4 & 52.7$\pm$0.7 & 11.2$\pm$0.3 & 28.3$\pm$0.6 & 3.1$\pm$0.6 & 51.2$\pm$1.8 &  9.8$\pm$0.7 & 27.1$\pm$1.4 & 4.8$\pm$0.9 & 54.1$\pm$1.2 & 13.6$\pm$0.6 & 31.5$\pm$1.1 & 1.0$\pm$0.3 & 51.1$\pm$1.5 & 5.0$\pm$0.4 & 22.3$\pm$0.8 & 1.1$\pm$0.3 & 51.5$\pm$0.4 & 5.2$\pm$0.2 & 22.1$\pm$0.3 &  7.80 \\
\midrule
KNN     & 3.9$\pm$0.2 & 52.4$\pm$0.2 & 10.5$\pm$0.5 & 27.8$\pm$0.7 & 3.8$\pm$0.5 & 52.1$\pm$1.8 & 10.9$\pm$0.6 & 28.2$\pm$1.1 & 4.2$\pm$0.9 & 53.8$\pm$1.0 & 12.8$\pm$0.5 & 30.1$\pm$1.1 & 1.1$\pm$0.4 & 51.5$\pm$1.3 & 5.1$\pm$0.3 & 22.5$\pm$0.9 & 1.1$\pm$0.3 & 51.4$\pm$0.6 & 4.9$\pm$0.2 & 21.0$\pm$0.1 &  8.18 \\
GRAPE   & 3.7$\pm$0.6 & 52.0$\pm$0.8 & 10.1$\pm$0.4 & 27.4$\pm$0.9 & 4.2$\pm$0.3 & 52.8$\pm$1.0 & 11.6$\pm$0.5 & 29.1$\pm$0.7 & 4.6$\pm$0.8 & 54.2$\pm$0.9 & 13.4$\pm$0.6 & 30.9$\pm$1.0 & 0.9$\pm$0.5 & 50.7$\pm$1.6 & 4.7$\pm$0.6 & 21.9$\pm$1.1 & 1.4$\pm$0.2 & 52.1$\pm$0.5 & 5.6$\pm$0.3 & 22.7$\pm$0.4 &  6.97 \\
VGAE    & 4.3$\pm$0.3 & 53.1$\pm$0.5 & 11.5$\pm$0.3 & 28.7$\pm$0.6 & 3.5$\pm$0.7 & 51.8$\pm$1.6 & 10.2$\pm$0.8 & 27.6$\pm$1.3 & 4.0$\pm$1.0 & 53.4$\pm$1.3 & 12.5$\pm$0.7 & 29.7$\pm$1.4 & \textbf{1.3$\pm$0.2} & 51.8$\pm$1.0 & 5.5$\pm$0.4 & 23.2$\pm$0.7 & 1.0$\pm$0.4 & 51.2$\pm$0.8 & 4.9$\pm$0.3 & 21.6$\pm$0.6 &  7.95 \\
\midrule
\qw{\texttt{LLM-L}}   & 4.9$\pm$0.0 & 53.0$\pm$0.4 & 13.2$\pm$0.2 & 29.9$\pm$0.2 & 9.2$\pm$0.9 & 65.1$\pm$1.5 & 16.2$\pm$0.9 & 31.6$\pm$0.4 & 20.2$\pm$1.6 & 72.1$\pm$0.5 & 29.2$\pm$1.7 & 43.7$\pm$0.9 & 1.2$\pm$0.3 & 51.0$\pm$1.3 & 6.0$\pm$0.9 & 22.4$\pm$0.3 & \underline{1.3$\pm$0.1} & 50.8$\pm$0.4 & 6.0$\pm$0.2 & 22.0$\pm$0.2 &  4.95 \\
\qw{\texttt{LLM-Q}}    & 5.6$\pm$0.3 & 54.1$\pm$0.6 & 14.7$\pm$0.4 & 31.2$\pm$0.5 & \underline{10.8$\pm$0.8} & 65.9$\pm$1.3 & \underline{17.8$\pm$0.9} & \underline{34.4$\pm$0.7} & 21.0$\pm$1.4 & 72.9$\pm$0.8 & \underline{31.2$\pm$1.6} & 44.8$\pm$1.2 & \textbf{1.7$\pm$0.3} & 51.6$\pm$1.2 & 6.5$\pm$0.5 & \underline{23.2$\pm$0.6} & 1.1$\pm$0.4 & 51.0$\pm$0.7 & 5.9$\pm$0.4 & 21.7$\pm$0.5 &  3.62 \\
\qw{\method-L$_I$} & \textbf{8.5$\pm$3.7} & \underline{54.7$\pm$2.5} & \underline{15.8$\pm$1.7} & \underline{32.0$\pm$1.4} & 10.5$\pm$0.8 & \underline{66.2$\pm$1.3} & 17.5$\pm$1.1 & 33.1$\pm$0.6 & \underline{21.3$\pm$1.8} & \underline{73.5$\pm$0.7} & 30.1$\pm$1.9 & \underline{45.2$\pm$1.2} & 1.4$\pm$0.2 & \textbf{51.8$\pm$1.5} & \underline{6.5$\pm$0.8} & 23.1$\pm$0.5 & \textbf{1.3$\pm$0.2 }& \textbf{54.2$\pm$0.2} & \textbf{6.7$\pm$0.1} & \textbf{22.8$\pm$0.2} &  \underline{2.23} \\
\qw{\method-Q$_I$}  & \underline{7.8$\pm$2.3} & \textbf{55.9$\pm$1.6 }& \textbf{17.2$\pm$1.4} & \textbf{33.4$\pm$1.2} & \textbf{11.6$\pm$0.9} & \textbf{67.1$\pm$1.5} & \textbf{18.9$\pm$1.0} & \textbf{35.2$\pm$0.8} & \textbf{22.5$\pm$1.7} & \textbf{74.8$\pm$0.9} & \textbf{32.8$\pm$1.8} & \textbf{46.9$\pm$1.3} & \underline{1.6$\pm$0.4} & 51.4$\pm$1.3 & \textbf{6.8$\pm$0.6} & \textbf{23.5$\pm$0.7} & 1.2$\pm$0.3 & \underline{53.9$\pm$0.6} & \underline{6.4$\pm$0.4} & \underline{22.6$\pm$0.4} & \textbf{ 1.77} \\
\bottomrule
\end{tabular}
\end{adjustbox}
\label{tab:rec_cold_start}
\end{table*}

\begin{table*}[t]
\centering
\footnotesize    
\setlength{\tabcolsep}{3pt}
\caption{Recommendation performance under the \textit{Uniform Masking} setting. LightGCN is an interaction-only baseline evaluated with the TWISTER-compatible in-batch negative protocol.}
\vspace{-1ex}
\label{tab:lightgcn_twister_uniform}
\begin{tabular}{lcccccccccccccccccccc}
\toprule
\multirow{2}{*}{Method}
& \multicolumn{4}{c}{Amazon\_Video}
& \multicolumn{4}{c}{Amazon\_Music}
& \multicolumn{4}{c}{Amazon\_Toys}
& \multicolumn{4}{c}{Goodreads\_Comics}
& \multicolumn{4}{c}{Goodreads\_Children} \\
\cmidrule(lr){2-5}
\cmidrule(lr){6-9}
\cmidrule(lr){10-13}
\cmidrule(lr){14-17}
\cmidrule(lr){18-21}
& ACC & AUC & MRR & NDCG
& ACC & AUC & MRR & NDCG
& ACC & AUC & MRR & NDCG
& ACC & AUC & MRR & NDCG
& ACC & AUC & MRR & NDCG \\
\midrule
LightGCN
& 1.8 & 56.4 & 7.5 & 23.4
& 0.5 & 51.5 & 5.4 & 21.6
& 2.9 & 51.0 & 9.2 & 25.1
& 1.8 & \textbf{59.2} & \textbf{7.4} & 23.6
& 0.8 & \textbf{58.3} & 6.3 & 22.5 \\
TWISTER-L
& 3.7 & 53.7 & 11.7 & \textbf{28.8}
& 7.0 & 68.5 & 17.1 & 32.6
& \textbf{28.9} & \textbf{80.7} & \textbf{40.2} & \textbf{51.9}
& 1.8 & 52.8 & 6.9 & 23.1
& \textbf{2.3} & 56.2 & 7.1 & 23.8 \\
TWISTER-Q
& \textbf{5.9} & \textbf{59.7} & \textbf{12.5} & 27.9
& \textbf{9.6} & \textbf{69.6} & \textbf{19.0} & \textbf{34.1}
& 6.2 & 59.1 & 19.1 & 34.9
& \textbf{2.6} & 54.2 & 7.3 & \textbf{24.1}
& 2.0 & 53.8 & \textbf{7.5} & \textbf{24.7} \\
\bottomrule
\end{tabular}
\end{table*}

\begin{table}[H]
  \caption{Amazon Product Review Evaluation}
  \vspace{-1em}
  \centering
  \footnotesize   
  \label{tab:amazon_prompt}

  \begin{tabular}{
    @{}>{\RaggedRight\arraybackslash}p{0.48\textwidth}@{}
  }
    \toprule
    \textbf{Amazon Product Review Evaluation Prompt} \\ \midrule
    You are evaluating the quality of an Amazon product review for a
    \texttt{\{category\}}. \\[0.2em]
    \textbf{Product ASIN:} \texttt{\{book\_id\}}\\
    \textbf{User ID:}      \texttt{\{user\_id\}}\\
    \textbf{Rating:}       \texttt{\{rating\}/5.0}\\
    \textbf{Review Text:}  ``\texttt{\{review\_text\}}''\\[0.2em]

    Please rate the review on a 5-point scale (1 = very poor, 5 = excellent):\\[-0.4em]
    \begin{itemize}[leftmargin=1em,itemsep=0pt,topsep=1pt]
        \item \textbf{Authenticity} – genuine, human-written tone.
        \item \textbf{Helpfulness} – useful to potential buyers.
        \item \textbf{Specificity} – concrete product details.
        \item \textbf{Readability} – clear and coherent.
    \end{itemize}

    \textbf{Important guidelines:}\\[-0.4em]
    \begin{itemize}[leftmargin=1em,itemsep=0pt,topsep=1pt]
        \item Empty or very short reviews should score lower.
        \item Generic or superlative-only language lowers authenticity.
        \item Consider whether the review is actionable.
        \item Judge coherence and grammar quality.
    \end{itemize}\\
    \textbf{Provide your evaluation in JSON below.}
    
\{ \\
   \texttt{$\quad$ "authenticity": <1-5>,} \\
    \texttt{$\quad$ "helpfulness":  <1-5>,}\\
  \texttt{$\quad$ "specificity":  <1-5>, }\\
 \texttt{$\quad$  "readability":  <1-5>,} \\
  \texttt{$\quad$ "reasoning":    "<brief explanation>" }\\
\} \\
\\
    \bottomrule
  \end{tabular}
  \vspace{-2ex}
\end{table}

\subsection{Capacity Lower Bound}
\begin{lemma}[Feature collapse from over-smoothing]
If $E(\mathbf{Z}) \to 0$, then for all $e, e'$, $\mathbf{z}_e \approx \mathbf{z}_{e'}$, so the prediction $\hat{y}_e$ becomes nearly constant. Thus, the model cannot fit variations in $y_e$ beyond the global mean, and $\mathcal{R}(\mathbf{w})$ is bounded below by the variance of the ratings.
\end{lemma}

\paragraph{Proof sketch.} Follows from the definition of Dirichlet energy: $E(\mathbf{Z}) = 0$ only if all $\mathbf{z}_e$ are identical (by Laplacian properties), so only the intercept can be fit.

\section{LLM-as-Judge Evaluation Prompt}
\label{sec:llm_judge_prompt}
Table~\ref{tab:amazon_prompt} provides the prompt used for
our LLM-as-Judge evaluation across Authenticity, Helpfulness,
Specificity, and Readability.

\section{Additional Recommendation Results}
\label{sec:supp_additional_rec}

\subsection{Full Cold-Start Results}
\label{sec:supp_cold_start}
Figure~\ref{fig:cold-start} in the main paper reports MRR and NDCG
under the Cold-Start setting for readability. Table~\ref{tab:rec_cold_start}
provides the corresponding complete results across ACC, AUC, MRR,
and NDCG for all five datasets and all baselines. The full results
follow the same pattern as Figure~\ref{fig:cold-start}:
the item-side TWISTER variants achieve the strongest average
performance when user-side review context is unavailable.

\subsection{Comparison with Interaction-Only Recommendation}
\label{sec:supp_lightgcn}
We additionally compare TWISTER with LightGCN, an interaction-only
recommender that does not rely on textual reviews. To ensure a
compatible comparison, LightGCN is evaluated using the same
in-batch negative protocol used in our recommendation experiments. As shown in Table~\ref{tab:lightgcn_twister_uniform}, TWISTER generally
provides substantially higher MRR and NDCG, supporting the utility
of recovering textual review signals rather than relying on
interaction structure alone.

\subsection{Alternative Downstream Backbone}
\label{sec:supp_gine}
To assess whether the benefit of TWISTER-imputed reviews depends
on the Edgeformers downstream architecture used in the main
experiments, we additionally evaluate GINE as an alternative
review-aware recommender. Table~\ref{tab:gine_results} reports
results under both Uniform Masking and Cold Start. Under Uniform
Masking, TWISTER-based review completion generally remains
competitive with or improves over zero-filled review features. The
benefit is clearer under Cold Start, where GINE with TWISTER
item-side imputation improves MRR and NDCG over both GINE-Rating,
GINE-ZeroFill, and structure-free LLM imputation. These results
indicate that the utility of TWISTER-imputed reviews is not specific
to the Edgeformers backbone.

\begin{table}[H]
\centering
\footnotesize
\setlength{\tabcolsep}{4pt}
\renewcommand{\arraystretch}{0.88}
\caption{GINE recommendation performance with an alternative downstream
backbone. Results are mean $\pm$ std over seeds, multiplied by 100.
Best results within each dataset and metric are bolded.}
\label{tab:gine_results}

\begin{tabular}{llcccc}
\toprule
Dataset & Variant & ACC & AUC & MRR & NDCG \\
\midrule

\multicolumn{6}{c}{\textit{Uniform Masking}} \\
\midrule
\multirow{4}{*}{Comics}
& GINE-Rating    & $1.5 \pm 0.4$ & $54.7 \pm 2.9$ & $6.6 \pm 0.8$ & $22.5 \pm 0.9$ \\
& GINE-ZeroFill  & $1.3 \pm 0.8$ & $60.1 \pm 1.5$ & $7.3 \pm 0.9$ & $23.5 \pm 0.8$ \\
& GINE-\method-L & $\mathbf{1.6 \pm 0.3}$ & $\mathbf{60.2 \pm 0.8}$ &
$\mathbf{7.6 \pm 0.4}$ & $\mathbf{23.8 \pm 0.4}$ \\
& GINE-\method-Q & $\mathbf{1.6 \pm 0.5}$ & $59.7 \pm 0.4$ &
$7.5 \pm 0.7$ & $23.6 \pm 0.6$ \\
\midrule

\multirow{4}{*}{Children}
& GINE-Rating    & $1.6 \pm 1.1$ & $56.2 \pm 1.6$ & $7.4 \pm 0.8$ & $23.5 \pm 0.7$ \\
& GINE-ZeroFill  & $2.6 \pm 0.6$ & $\mathbf{59.7 \pm 1.6}$ & $8.8 \pm 0.4$ & $24.9 \pm 0.4$ \\
& GINE-\method-L & $2.2 \pm 0.6$ & $59.5 \pm 1.9$ & $8.5 \pm 0.7$ & $24.7 \pm 0.6$ \\
& GINE-\method-Q & $\mathbf{2.8 \pm 0.8}$ & $\mathbf{59.7 \pm 1.9}$ &
$\mathbf{9.3 \pm 0.8}$ & $\mathbf{25.3 \pm 0.7}$ \\

\midrule
\multicolumn{6}{c}{\textit{Cold Start}} \\
\midrule

\multirow{4}{*}{Comics}
& GINE-Rating & $1.3 \pm 0.4$ & $55.6 \pm 1.9$ & $6.5 \pm 0.3$ & $22.5 \pm 0.4$ \\
& GINE-ZeroFill & $1.2 \pm 0.5$ & $61.7 \pm 1.0$ & $7.7 \pm 0.8$ & $24.0 \pm 0.7$ \\
& GINE-\method-I & $\mathbf{1.8 \pm 0.8}$ & $\mathbf{63.1 \pm 1.0}$ &
$\mathbf{8.3 \pm 0.8}$ & $\mathbf{24.6 \pm 0.7}$ \\
& GINE-LLM & $1.2 \pm 0.7$ & $59.4 \pm 1.0$ & $7.3 \pm 0.4$ & $23.4 \pm 0.3$ \\
\midrule

\multirow{4}{*}{Children}
& GINE-Rating & $1.7 \pm 0.3$ & $56.5 \pm 1.4$ & $7.4 \pm 0.3$ & $23.5 \pm 0.3$ \\
& GINE-ZeroFill & $2.0 \pm 0.6$ & $60.1 \pm 1.5$ & $8.3 \pm 0.6$ & $24.5 \pm 0.6$ \\
& GINE-\method-I & $\mathbf{2.2 \pm 0.7}$ & $\mathbf{60.4 \pm 2.0}$ &
$\mathbf{8.9 \pm 0.8}$ & $\mathbf{25.0 \pm 0.8}$ \\
& GINE-LLM & $1.6 \pm 0.7$ & $60.2 \pm 1.5$ & $8.2 \pm 0.8$ & $24.5 \pm 0.7$ \\

\bottomrule
\end{tabular}
\end{table}


\end{document}